\documentclass[12pt, peerreview, onecolumn]{IEEEtran}
\pdfoutput=1
\usepackage{amsmath,epsfig,amssymb,verbatim,amsopn,subfigure,color}
\usepackage{cite,xspace}
\usepackage{array,algorithm,algorithmic}

\ifCLASSOPTIONpeerreview
    \usepackage[nomarkers]{endfloat}
\fi

\ifCLASSOPTIONonecolumn
\AtBeginDelayedFloats{} 
\fi

\ifCLASSOPTIONpeerreview

\fi

\ifCLASSOPTIONonecolumn

\else
    
\fi

\usepackage{url}
\usepackage{ifpdf}
\ifpdf
\usepackage[pdftex]{hyperref}
\else
\usepackage[ps2pdf]{hyperref}
\fi

\makeatletter
\renewcommand*{\@opargbegintheorem}[3]{\trivlist
  \item[\hskip \labelsep{\itshape #1\ #2}] {\itshape (#3):} {\normalfont}}
\makeatother

\newcommand{\dv}[1]{\boldsymbol{#1}}
\newcommand{\pga}{\mathcal{A}}
\newcommand{\cda}{\mathcal{D}}

\newcommand{\rop}{\mathfrak{R}}
\newcommand{\bbb}{\mathcal{B}}
\newcommand{\ccc}{\mathcal{C}}

\newcommand{\rr}{r}
\newcommand{\el}{L}

\newcommand{\figref}[1]{Fig.\,\ref{#1}}

\newcommand{\appref}[1]{Appendix\,\ref{#1}}

\usepackage{relsize}

\newtheorem{definition}{Definition}

\graphicspath{{figs/},{Figures/}}

\newcommand{\AuthorOne}{Zubair~Khalid, {\em{Student Member, IEEE}}}
\newcommand{\AuthorTwo}{Salman~Durrani, {\em{Senior Member, IEEE}}}

\newcommand{\ThankOne}{The authors are with the Research School
of Engineering, College of Engineering and Computer Science, The
Australian National University, Canberra, ACT 0200, Australia.
Emails: \{zubair.khalid, salman.durrani\}@anu.edu.au}

\begin{document}

\title{{Distance Distributions in Regular Polygons}}
\author{\authorblockN{\AuthorOne \: and \AuthorTwo
\thanks{\ThankOne}}}

\maketitle

%
\begin{abstract}
This paper derives the exact cumulative density function of the distance between a randomly located node and any arbitrary reference point inside a regular $\el$-sided polygon. Using this result, we obtain the closed-form probability density function (PDF) of the Euclidean
distance between any arbitrary reference point and its $n$-th
neighbour node, when $N$ nodes are uniformly and independently
distributed inside a regular $\el$-sided polygon. First, we exploit the rotational symmetry
of the regular polygons and quantify the effect of polygon sides and
vertices on the distance distributions. Then we propose an algorithm to
determine the distance distributions given any arbitrary location of the reference point inside the polygon. For the special case when the
arbitrary reference point is located at the center of the polygon,
our framework reproduces the existing result in the literature.

\end{abstract}

\ifCLASSOPTIONpeerreview
    \vspace{-0.15in}
\fi


\ifCLASSOPTIONpeerreview
    \vspace{-0.15in}
\fi

\section{Introduction}
Recently, the distance distributions in wireless networks have
received a lot of attention in the
literature~\cite{Srinivasa-2010,Moltchanov-2012,Fan-2007}. The
distance distributions can be applied to study important wireless
network characteristics such as interference, outage probability,
connectivity, routing and energy consumption~\cite{Srinivasa-2010,Haenggi-2008,Andrews-2012,Fabbri-08}.
The distance distributions in wireless networks are dependent on the
location of the nodes, which are seen as realizations of some
spatial point process. When the node locations follows an infinite
homogeneous Poisson point process, the probability density function
(PDF) of the Euclidean distance between a point and its $n$-th
neighbour node follows a generalised Gamma
distribution~\cite{Haenggi-2005}. However, as recently identified
in~\cite{Srinivasa-2010,Moltchanov-2012,Torrieri-2012,Mao-2012}, this model does not accurately reflect
the distance distributions in many practical wireless networks where
a fixed and finite number of nodes are uniformly and independently
distributed over a finite area such as square or hexagon or disk
region. Note that these finite regions of interest can be
conveniently modeled as special cases of a regular $\el$-sided
polygon (referred to as $\el$-gon for brevity), e.g.,
$\el=3,4,6,\infty$ correspond to equilateral triangle, square,
hexagon and disk respectively. In this context, the two important
distance distributions are: (i) the PDF of the Euclidean distance
between two nodes uniformly and independently distributed inside a
$\el$-gon and (ii) the PDF of the Euclidean distance between any
arbitrary reference point and its $n$-th neighbour node, when $N$
nodes are uniformly and independently distributed inside a
$\el$-gon. For the first case, the PDF of the distance between two
nodes uniformly and independently distributed inside an equilateral
triangle~\cite{Zhuang-2012}, square~\cite{Mathai-1999,Kostin-2011},
hexagon~\cite{Zhuang-2011b,Fan-2007} and disk~\cite{Mathai-1999} are
well known in the literature. These results are special
cases of the general result obtained recently in~\cite{Basel-2012}.
For the second case, the PDF of the Euclidean distance to the $n$-th
neighbour node is obtained in~\cite{Srinivasa-2010} for the special
case when the reference point is located at the center of the
$\el$-gon.

In this correspondence, we present a general framework for analytically obtaining the exact cumulative density function of the distance between a randomly located node and any arbitrary reference point inside a regular $\el$-sided polygon. Using this result, we obtain the closed-form PDF of the Euclidean distance between any arbitrary reference point and its $n$-th neighbour node, when $N$ nodes are uniformly and independently distributed inside a regular $\el$-sided polygon. The proposed framework is based on characterising the overlap area between the $\el$-gon and a disk centered at the arbitrary reference point located inside the $\el$-gon. There are two key insights which lead to our results: the use of the rotation operator that simplifies the characterisation of distances and overlap areas, and the systematic analysis of the effect of $\el$-gon sides and vertices on the overlap area. Based on our proposed framework, we formulate an algorithm to determine the distance distributions given any location of the reference point inside the polygon. We provide examples to demonstrate the generality of our proposed framework. We also show that the result in~\cite{Srinivasa-2010} can be obtained as a special case in our framework.

\ifCLASSOPTIONpeerreview
    \vspace{-0.15in}
\fi

\section{System Model}\label{sec:sysmodel}

Consider $N$ nodes which are uniformly and independently distributed
inside a regular $\el$-sided polygon $\mathcal{A}\in \mathbb{R}^2$,
where $\mathbb{R}^2$ denotes the two dimensional Euclidean domain.
Let $\dv{u}=[x,\, y]^{T}\in\mathcal{A}$ denote an arbitrary
reference point located inside the $\el$-gon, where $[\cdot]^T$ denotes transpose of a vector.

\ifCLASSOPTIONonecolumn
\begin{figure*}[t]
 \centering
 \hspace{-18mm}
    \includegraphics[width=0.6\textwidth]{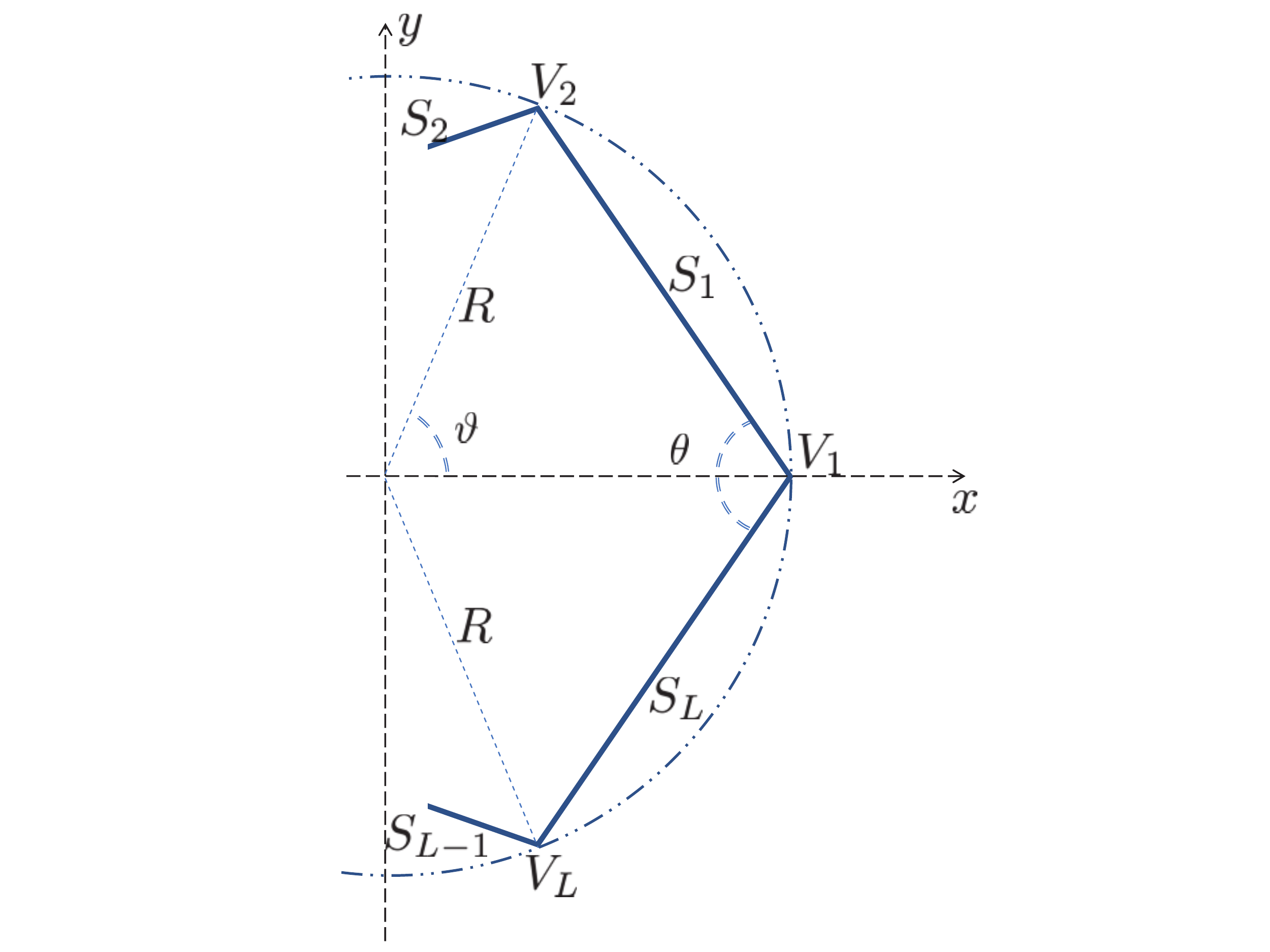}
    \vspace{-0mm} \hspace{-15mm}

\vspace{-1mm} \caption{Illustration of sides, vertices and angles
for the $L$-sided polygon inscribed in a circle of radius $R$.
$\theta$ denoting the interior angle of the polygon and $\vartheta$
denoting the central angle between two adjacent vertices are defined
in \eqref{eq:angles}.} \label{fig:one}
\end{figure*}
\else
\begin{figure}[t]
 \centering
 \hspace{-22mm}
    \includegraphics[width=0.55\textwidth]{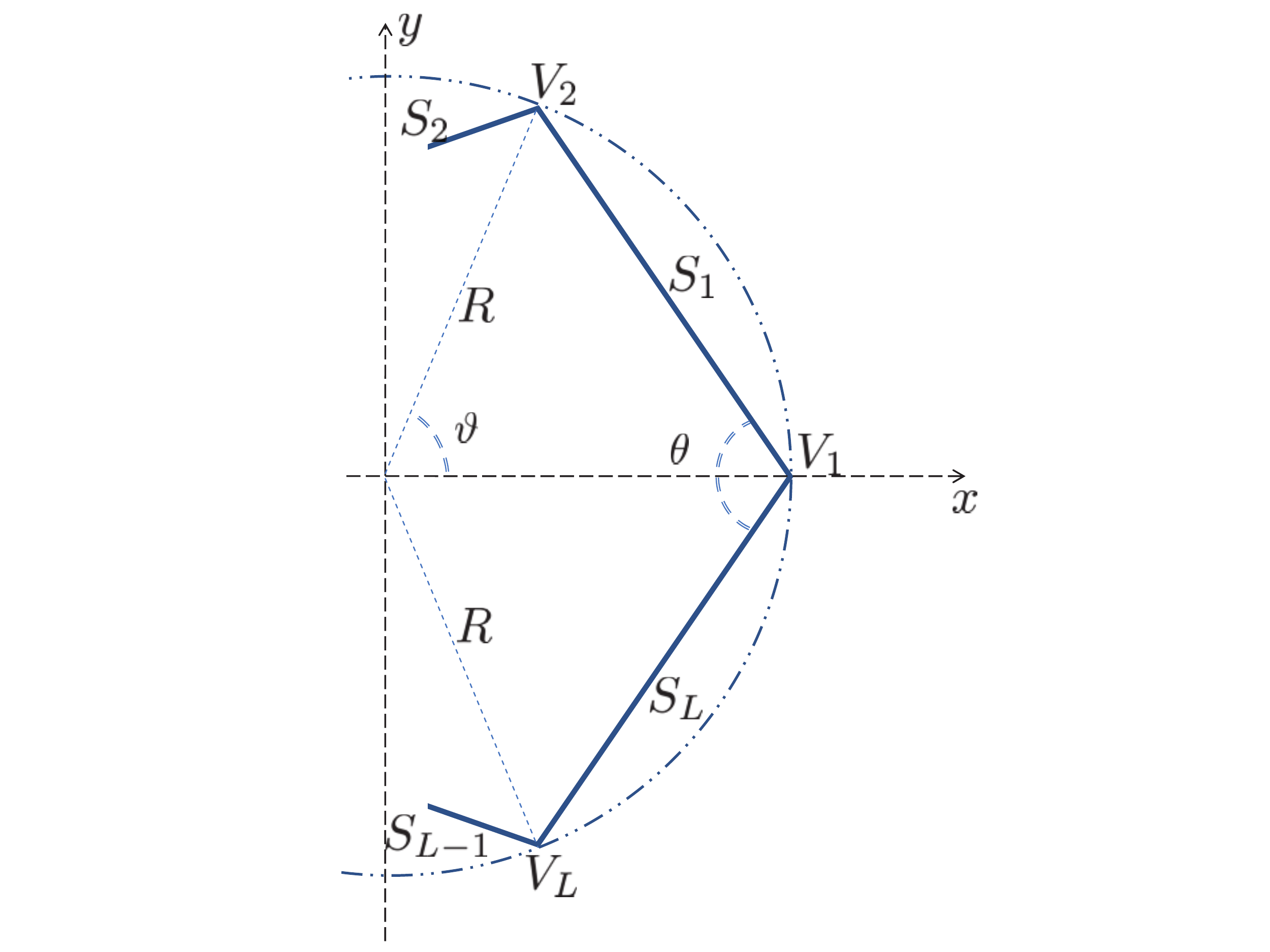}
    \vspace{-0mm} \hspace{-15mm}

\vspace{-1mm} \caption{Illustration of sides, vertices and angles
for the $L$-sided polygon inscribed in a circle of radius $R$.
$\theta$ denoting the interior angle of the polygon and $\vartheta$
denoting the central angle between two adjacent vertices are defined
in \eqref{eq:angles}.} \label{fig:one}
\end{figure}
\fi

\ifCLASSOPTIONpeerreview
    \vspace{-0.15in}
\fi
\subsection{Polygon geometry}
Without loss of generality, we assume that the $\el$-gon is
inscribed in a circle of radius $R$ and is centered at the origin
$[0,\, 0]^T$. Then, its inradius is $R_{\textrm{i}} = R \cos
(\pi/\ell)$ and its area $A$ is given by
\begin{align}\label{eq:area}
A = |\mathcal{A}|= \frac{1}{2}\el R^2\sin
\left(\frac{2\pi}{\el}\right).
\end{align}

Let $S_\ell$ and $V_\ell$ denotes the sides and vertices of the
polygon, for $\ell=1,2,\hdots,L$, which are numbered in
anti-clockwise direction, as shown in \figref{fig:one}. We assume
that the first vertex $V_1$ of the polygon is at $[R,\, 0]^T$, i.e.,
at the intersection of the circle inscribing the polygon and the
$x$-axis. The interior angle of the polygon $\theta$ and the central
angle between two adjacent vertices $\vartheta$ are given by
\begin{align}\label{eq:angles}
\theta &=\frac{\pi(\el-2)}{\el} \qquad \textrm{and}\qquad \vartheta =\frac{2\pi}{\el}. %
\end{align}

\ifCLASSOPTIONpeerreview
    \vspace{-0.25in}
\fi

\subsection{Rotation operator} \label{sec:rot}
For compact representation, we define the rotation operator
$\rop^\ell$ which rotates an arbitrary point $\dv{u}=[x,\, y]^T$
anti-clockwise around the origin by an angle $\ell \vartheta$. The
rotated point $\rop^\ell \dv{u}$ can thus be expressed as
\begin{align}\label{eq:rot}
(\rop^\ell \dv{u}) = \mathbf{T}\,\dv{u},
\end{align}

\noindent where $\mathbf{T}$ is the corresponding rotation matrix of
$\rop^\ell$ and is given by
\begin{align}
\mathbf{T}\,=\, \left( \begin{array}{cc}
\cos(\ell\vartheta) & -\sin(\ell\vartheta)  \\
\sin(\ell\vartheta) & \cos(\ell\vartheta)  \end{array} \right).
\end{align}

\noindent Also define $\rop^{-\ell}$ as the inverse rotation
operator with rotation matrix $\mathbf{T}^{-1}$, which rotates an
arbitrary point $\dv{u}=[x,\, y]^T$ clockwise around the origin by
an angle $\ell \vartheta$. We note that the rotation under the
operator $\rop$ is an isometric operation which preserves the
distances between any two points in the two dimensional plane, i.e.,
$\det(\mathbf{T})=1$ and $\mathbf{T}\mathbf{T}^{-1}=\mathbf{I}$,
where $\mathbf{I}$ denotes the identity matrix.

\ifCLASSOPTIONpeerreview
    \vspace{-0.15in}
\fi

\subsection{Distances from the polygon sides and vertices}

In this subsection, we find the distances from an arbitrary reference point $\dv{u}$ located inside the polygon to all its sides and vertices. First, we examine the distance to any vertex. Using the geometry, the
distance $d(\dv{u};\, V_{1})$ between the point $\dv{u}$ and the vertex $V_1$ is given by
\begin{align}
d(\dv{u};\, V_{1}) \,&=\, \sqrt{(x-R)^2+y^2}.
\label{Eq:distances_first2}
\end{align}

In order to find the distances to the remaining vertices, we use $d(\dv{u};\, V_{1})$ and exploit the rotational symmetry of the $\el$-gon. By appropriately rotating the point $\dv{u}$ and then finding its distance from vertex $V_1$, we can in fact, find the distance to other vertices. Thus, using~(\ref{Eq:distances_first2}) and the rotation operator defined in \eqref{eq:rot}, we can express $d(\dv{u};\, V_{\ell})$ as
\begin{align}
d(\dv{u};\, V_{\ell}) \,&=\, d(\rop^{-(\ell-1)}\dv{u};\,V_{1}).\label{Eq:distances_general2}
\end{align}

Next, we examine the distance to any side. We define this distance to any side as the \textit{shortest} distance between the point $\dv{u}$ and the line segment formed by the side of the polygon. If the projection of the point $\dv{u}$ onto a side lies inside the side, then the shortest distance is given by the perpendicular distance between the point and the side. If the projection of the point $\dv{u}$ onto a side does not lie on the side, then the shortest distance is the minimum of the distance to the side endpoints. Using the geometry, the perpendicular distance from an arbitrary point $\dv{u}$ to the line segment formed by side $S_1$ is given by
\begin{align}
 p(\dv{u};\,S_1) \,&=\, \frac{\textrm{abs}\left(y + \tan
\left(\frac{\theta}{2}\right)\,x - R \tan \left(\frac{\theta}{2}\right)\right)}{\sqrt{ 1 +
\tan^2  \left(\frac{\theta}{2}\right)} }, \label{Eq:p_distances_first1}
\end{align}

\noindent where $\mathrm{abs}(\cdot)$ denotes the absolute value. The distance to the side $S_1$ endpoints is simply given by $d(\dv{u};\, V_{1})$ and $d(\dv{u};\, V_{2})$. Thus, the shortest distance $d(\dv{u};\,S_1)$ to the side $S_1$ can be expressed as
\ifCLASSOPTIONonecolumn
\begin{align}\label{Eq:distances_first1}
d(\dv{u};\,S_1)\,&=\,
\begin{cases}
\min(d(\dv{u};\, V_{1}),\,d(\dv{u};\, V_{2})), & \max(d(\dv{w}; V_{1}),\,d(\dv{w}; V_{2}))>t ;\\
p(\dv{u};\,S_1), & \textrm{otherwise};
\end{cases}
\end{align}
\else
\begin{align}\label{Eq:distances_first1}
d(\dv{u};\,S_1)\,&=\,
\begin{cases}
\min(d(\dv{u};\, V_{1}),\,d(\dv{u};\, V_{2})), & \\ \quad \quad  \quad \quad \quad \max(d(\dv{u}_1; V_{1}),\,d(\dv{u}_1; V_{2}))>t;\\
p(\dv{u};\,S_1), & \hspace{-47mm} \textrm{otherwise};
\end{cases}
\end{align}
\fi

\noindent where $\dv{w} = \big[R - \frac{(x-R)(\cos\vartheta-1)
+ y\sin\vartheta}{2} ,\, \frac{\sin\vartheta\big( (x-R)(\cos\vartheta-1)
+ y \sin\vartheta \big) }{2(1 - \cos\vartheta)}\big]^T$ denotes the perpendicular projection of
$\dv{u}$ onto the line segment formed by side $S_1$,
$t=2R\sin\left(\frac{\pi}{L}\right)$ is the side length of the
$L$-gon and $\min(\cdot)$ and $\max(\cdot)$ denote the
minimum and maximum values respectively. Then, using \eqref{eq:rot}, we can express $p(\dv{u};\,
S_{\ell})$ and $d(\dv{u};\, S_{\ell})$ as
\begin{align}
p(\dv{u};\,S_\ell)\,=\, p(\rop^{-(\ell-1)}\dv{u};\,S_1), \quad
d(\dv{u};\,S_\ell)\,=\,
d(\rop^{-(\ell-1)}\dv{u};\,S_1),\label{Eq:distances_general1}
\end{align}

\begin{definition}
Define the distance vector $\mathbf{d}$ as an indexed vector of the
distances from the arbitrary point $\dv{u}\in\pga$ to all the sides
and vertices of the $\el$-gon, defined in
\eqref{Eq:distances_first2}, \eqref{Eq:distances_general2}, \eqref{Eq:distances_first1}, and \eqref{Eq:distances_general1} as
\begin{align}\label{eq:distvector}
\mathbf{d}\,=\,[d(\dv{u};S_1),\,\hdots,\,d(\dv{u};S_L),\,d(\dv{u};V_1),\,\hdots,\,d(\dv{u};V_L)].
\end{align}
\end{definition}
\section{Problem Formulation}\label{sec:prob}

Consider a disk $\cda(\dv{u};\rr)$ of radius $r$ centered at the
arbitrary reference point $\dv{u}$. First, we define the probability
that a random node, which is uniformly distributed inside the
$\el$-gon $\mathcal{A}$, lies at a distance of less than or equal to
$\rr$ from the point $\dv{u}$.

\begin{definition}
Define the cumulative density function~(CDF) $F(\dv{u};\rr)$, which
is the probability that a random node falls inside a disk
$\cda(\dv{u};\rr)$ of radius $r$ centered at the arbitrary reference
point $\dv{u}$, as
\begin{align}\label{Eq:CDF_general}
F(\dv{u};\rr)\,&=\, \frac{| \cda(\dv{u};\rr)\cap\pga|}{|\pga|}= \frac{O(\dv{u};\rr)}{A}
\end{align}

\noindent where $O(\dv{u};\rr)=| \cda(\dv{u};\rr)\cap\pga|$ is the overlap area between the disk $\cda(\dv{u};\rr)$ and $\el$-gon $\mathcal{A}$.
\end{definition}

Then the PDF $f_n(\dv{u};\rr)$ of the distance from an arbitrary
point $\dv{u}$ to the $n$-th neighbour node is~\cite{Srinivasa-2010}
\begin{align}\label{Eq:pdf_general}
f_n(\dv{u};\rr)\,=\,
\frac{\big(1-F(\dv{u};\rr)\big)^{N-n}\big(F(\dv{u};\rr)\big)^{n-1}}{B(N-n+1,k)}
\,\frac{\partial\,F(\dv{u};\rr)}{\partial\rr}
\end{align}

\noindent where $N$ is the number nodes are which are uniformly and
independently distributed inside the $\el$-gon, $B(a,b) =
\Gamma(a)\Gamma(b)/\Gamma(a+b)$ is the beta function,
$\Gamma(\cdot)$ denotes the gamma function,
$\frac{\partial}{\partial\rr}(\cdot)$ denotes the partial derivative
respect to the variable $r$ and $F(\dv{u};\rr)$ is given
in~\eqref{Eq:CDF_general}.

The challenge in evaluating \eqref{Eq:CDF_general} and
\eqref{Eq:pdf_general} is finding the overlap area $O(\dv{u};\rr)$.
When the \textit{reference point is located at the center} of the
$\el$-gon, the overlap area $O(\dv{u};\rr)$ can be easily evaluated
as shown in~\cite{Srinivasa-2010}. For $0 \leq r \leq
R_{\textrm{i}}$ (where $R_{\textrm{i}}$ is the $\el$-gon inradius
defined above~\eqref{eq:area}), the disk $\cda(\dv{u};\rr)$ is
entirely inside the $\el$-gon $\mathcal{A}$. Thus the overlap area
is the same as the area of the disk, i.e, $O(\dv{u};\rr)=\pi r^2$.
For $R_{\textrm{i}} \leq r \leq R$, (where $R$ is the $\el$-gon
circumradius) there are $L$ circular segment shaped portions of the
disk $\cda(\dv{u};\rr)$ which are outside the $\el$-gon. Since these
circular segments are symmetric and there is no overlap between
them, the overlap area is given by $O(\dv{u};\rr)=\pi
r^2-L(\textrm{area of one circular segment})$. Note that in this
case it is simple to find the area of one circular segment using
geometry.

When the \textit{reference point is not located at the center}, then
the circular segments are no longer symmetric. This is illustrated
in \figref{three} for the case of a square ($\el=4$) for simplicity.
We can see that for the given radius of the disk $\cda(\dv{u};\rr)$
centered at the reference point, non-symmetric circular segment
areas are formed outside the sides $S_1$, $S_2$ and $S_4$. In
addition, there is \textit{overlap} between the circular segment
areas formed outside the sides $S_1$ and $S_2$. This complicates the
task of finding the overlap area $O(\dv{u};\rr)$. Also, for any
arbitrary location of the reference point, the different ranges for
$r$ are no longer exclusively determined by the inradius
$R_{\textrm{i}}$ and circumradius $R$, but by the number of unique
distances from the reference point to the vertices and the sides. This adds further
complexity to the task of finding the overlap area $O(\dv{u};\rr)$.

In the next section, we propose our method to systematically account
for the effect of sides and vertices on the overlap area
$O(\dv{u};\rr)$. Then in Section~\ref{sec:algo}, we propose an
algorithm to automatically formulate the different ranges for $\rr$
and find the overlap area $O(\dv{u};\rr)$ in order to evaluate
\eqref{Eq:CDF_general} and \eqref{Eq:pdf_general}.
%
\ifCLASSOPTIONonecolumn
\begin{figure*}[t]
 \centering
 \hspace{-18mm}
    \includegraphics[width=0.6\textwidth]{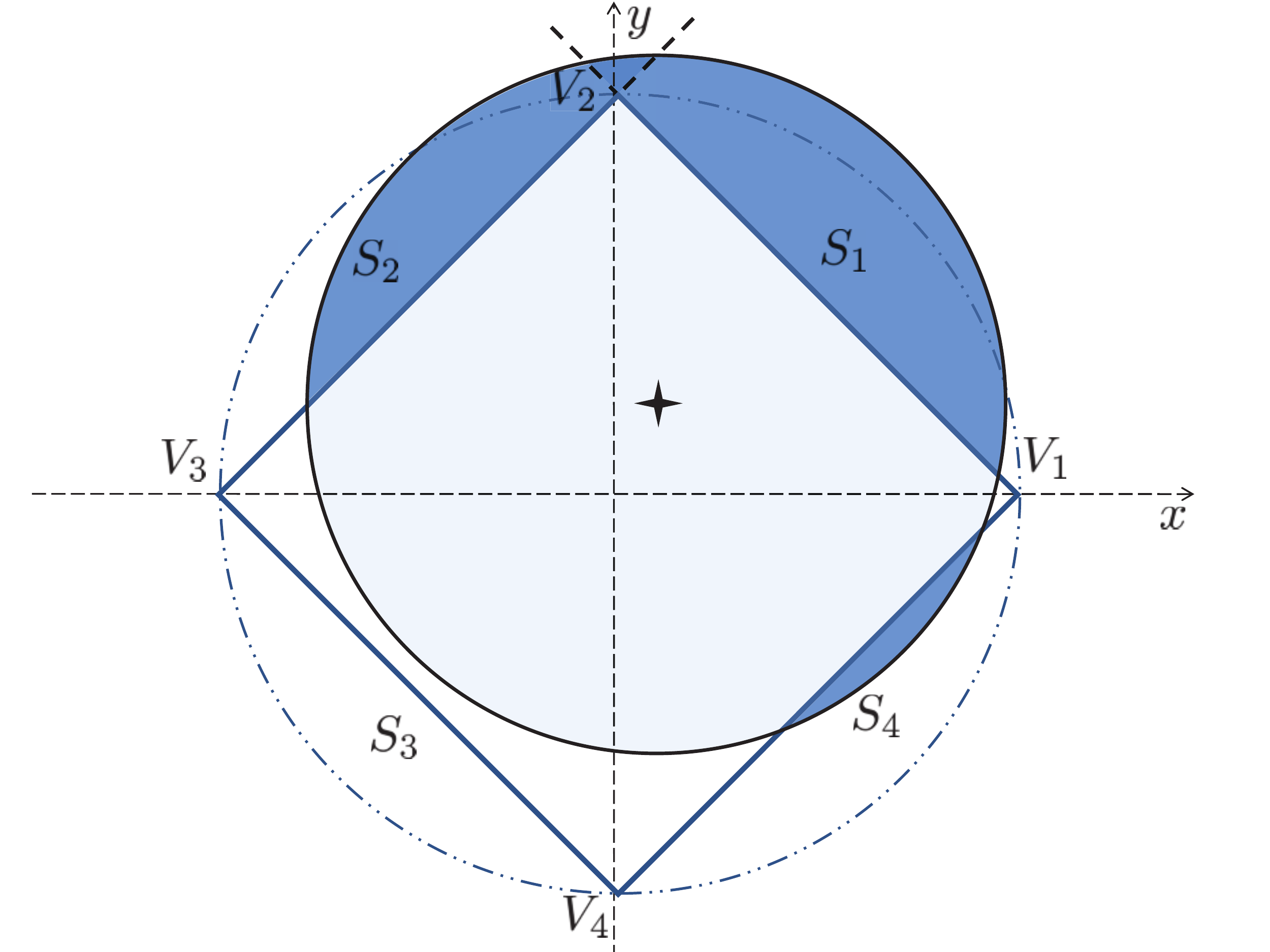}
    \vspace{-0mm} \hspace{-15mm}

\vspace{-1mm} \caption{Illustration of the non-symmetric circular
segment areas formed outside the sides $S_1$, $S_2$ and $S_4$ for a
square ($L=4$-gon). There is also an overlap between the circular
segment areas formed outside the sides $S_1$ and $S_2$.}
\label{three}
\end{figure*}
\else
\begin{figure}[t]
 \centering
    \includegraphics[width=0.52\textwidth]{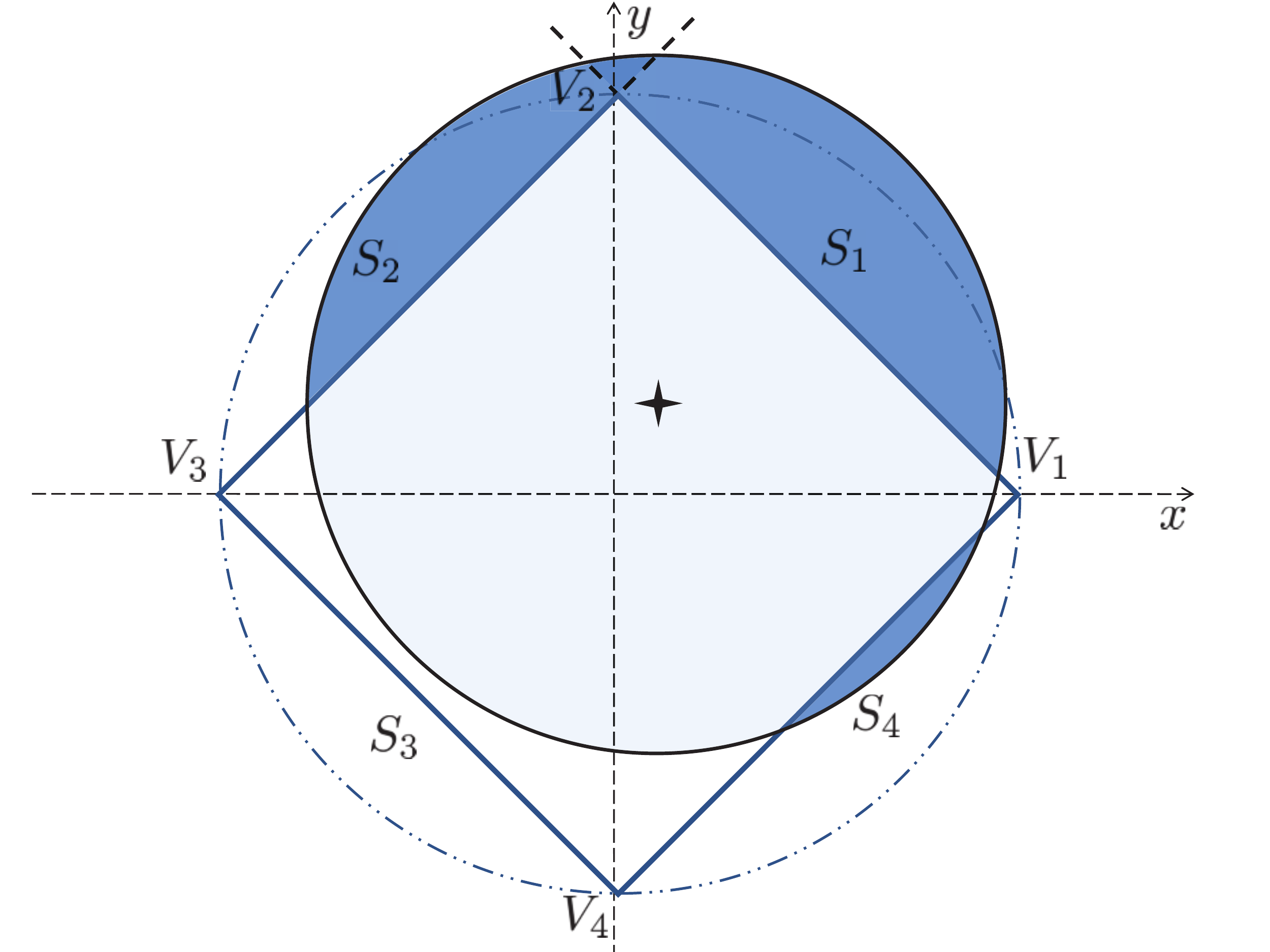}

\vspace{-2mm} \caption{Illustration of the non-symmetric circular
segment areas formed outside the sides $S_1$, $S_2$ and $S_4$ for a
square ($L=4$-gon). There is also an overlap between the circular
segment areas formed outside the sides $S_1$ and $S_2$.}
\label{three}
\end{figure}
\fi


\section{Characterization of the Effect of Sides and Vertices}
In this section, we characterize the effect of polygon sides and vertices on the overlap area $O(\dv{u};\rr)$. Because of
the symmetry of the polygon, we only need to consider the following
three cases which are illustrated in \figref{fig:two}. Other cases can be handled as an appropriate combination of these cases:
\begin{enumerate}
  \item The overlap area is limited by one side of the polygon only. This is illustrated
in \figref{fig:two}(a) for side $S_1$.
  \item The overlap area is limited by two sides of the polygon and there is no
overlap between them. This is illustrated in \figref{fig:two}(b) for
sides $S_1$ and $S_L$.
  \item The overlap area is limited by two sides of the polygon and there is overlap
between them. This is illustrated in \figref{fig:two}(c) for side $S_1$
and $S_L$ and vertex $V_1$.
\end{enumerate}

\noindent These cases are discussed in detail in the following subsections.

\ifCLASSOPTIONonecolumn
\begin{figure*}[t]
 \centering
 \subfigure[]
 {
    \includegraphics[width=0.5\textwidth]{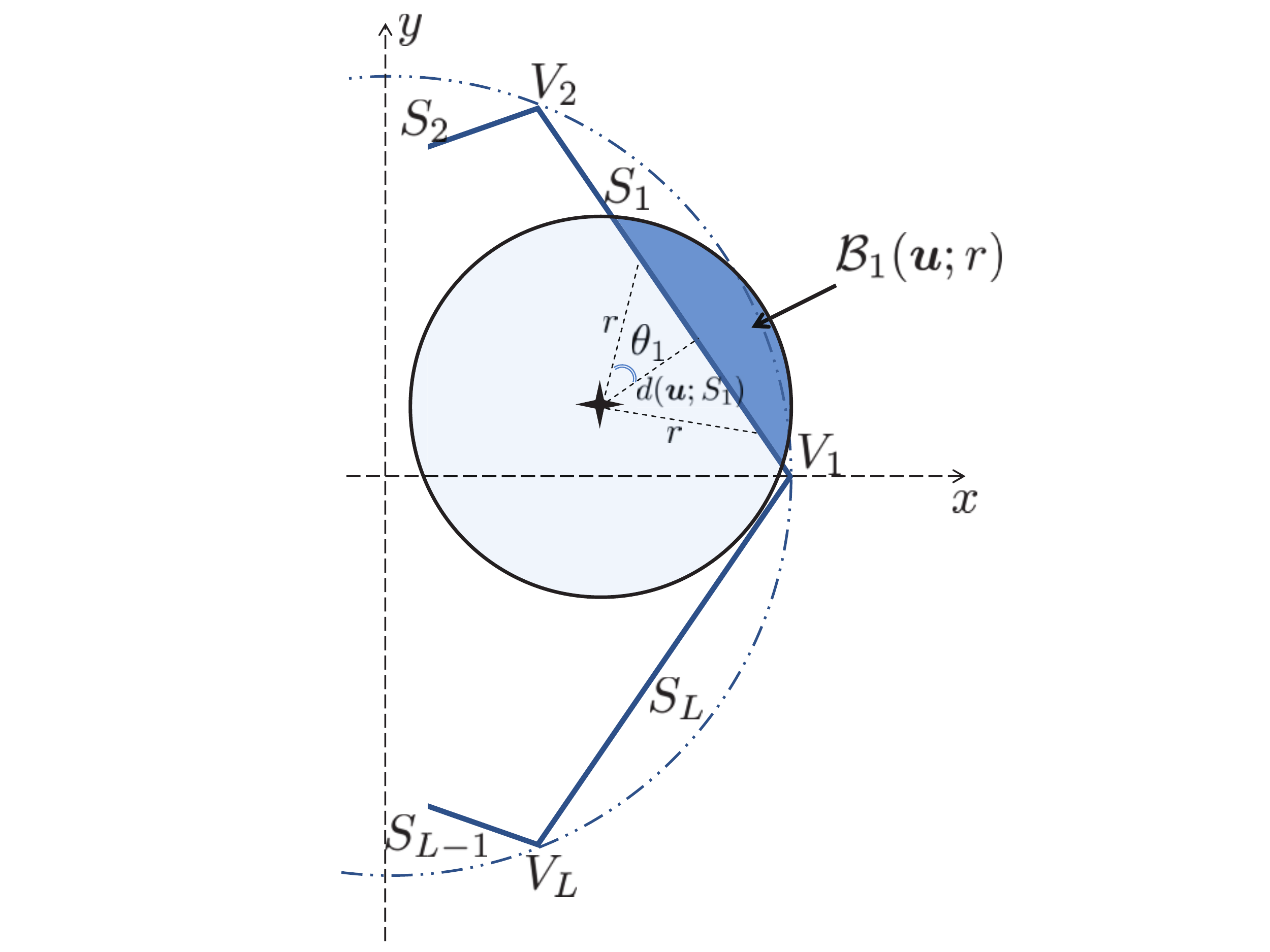}
    }\vspace{-0mm}

 \subfigure[]
 {
    \includegraphics[width=0.5\textwidth]{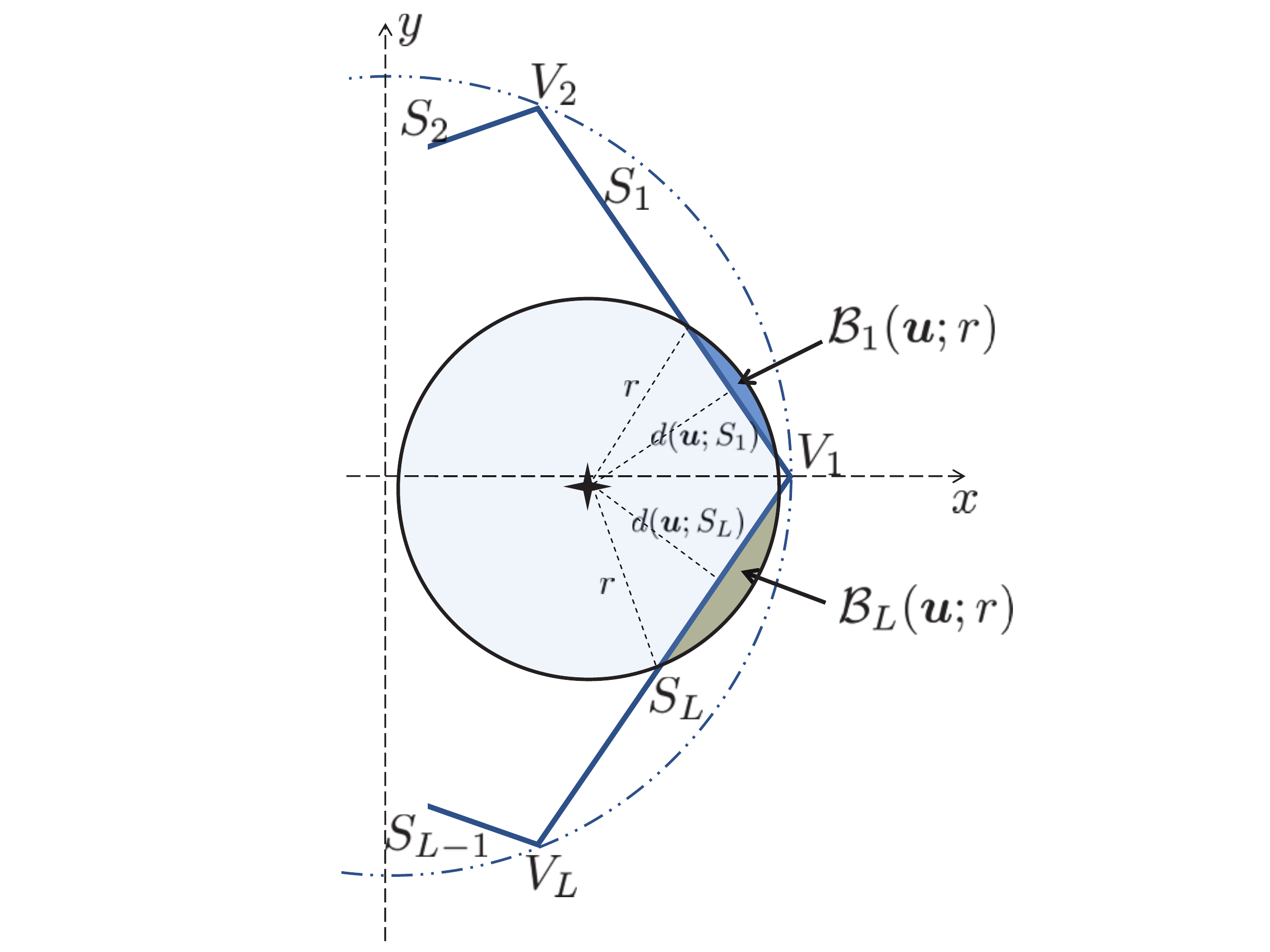}
 }\hspace{-8mm}

 \subfigure[]
 {
    \includegraphics[width=0.5\textwidth]{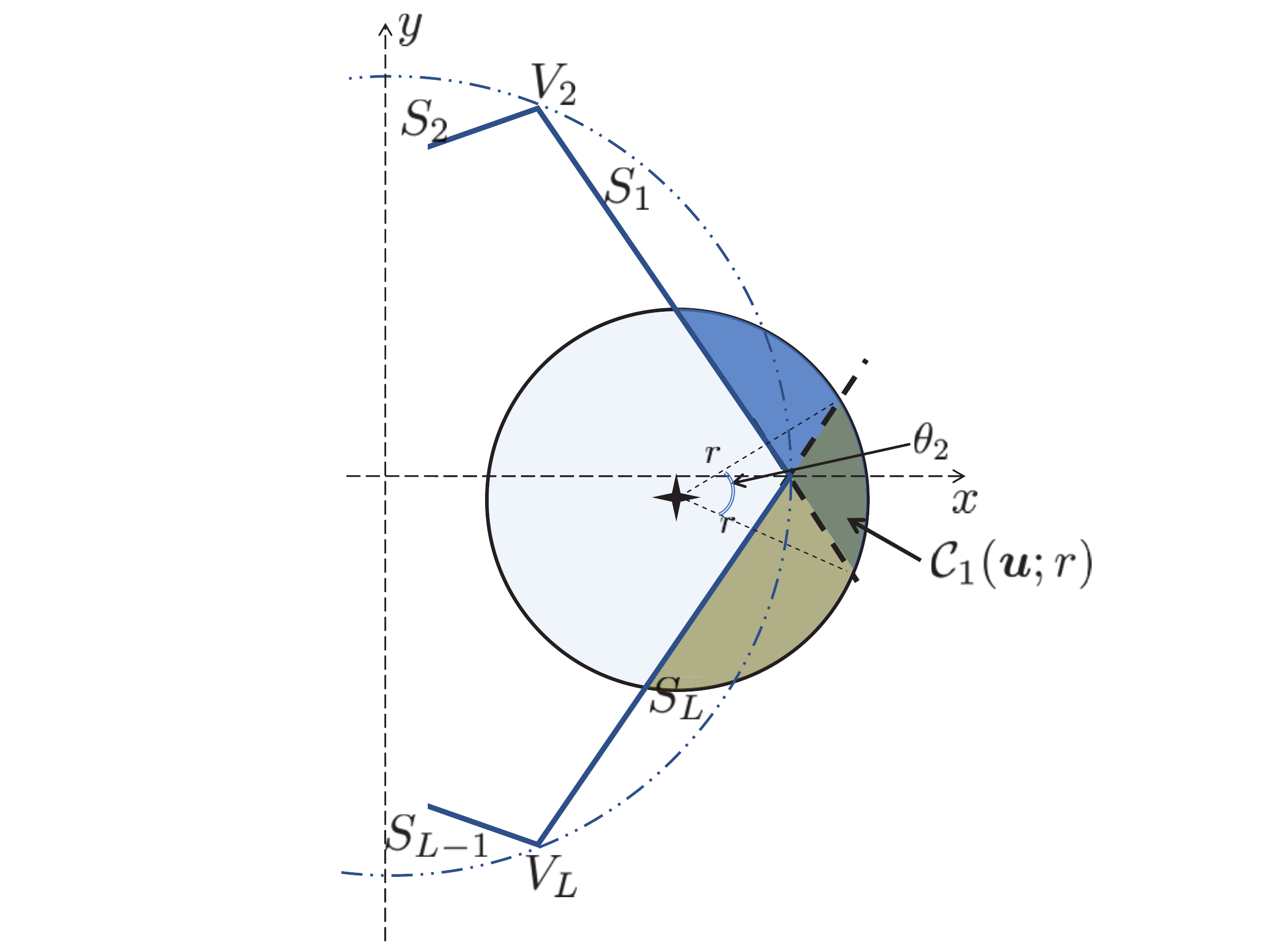}
 }\vspace{-0mm} \hspace{-4mm}
\vspace{-1mm} \caption{Illustration of the effect of sides and
vertices of the polygon on the overlap are: (a) overlap area is
limited by side $S_1$, (b) overlap area is limited by sides $S_1$
and $S_L$ only and (c) overlap area is limited by sides $S_1$ and
$S_L$ with inclusion of the vertex $V_1$.} \label{fig:two}
\end{figure*}
\else
\begin{figure*}[th]
 \hspace{-15mm}
 \subfigure[]
 {
    \includegraphics[width=0.43\textwidth]{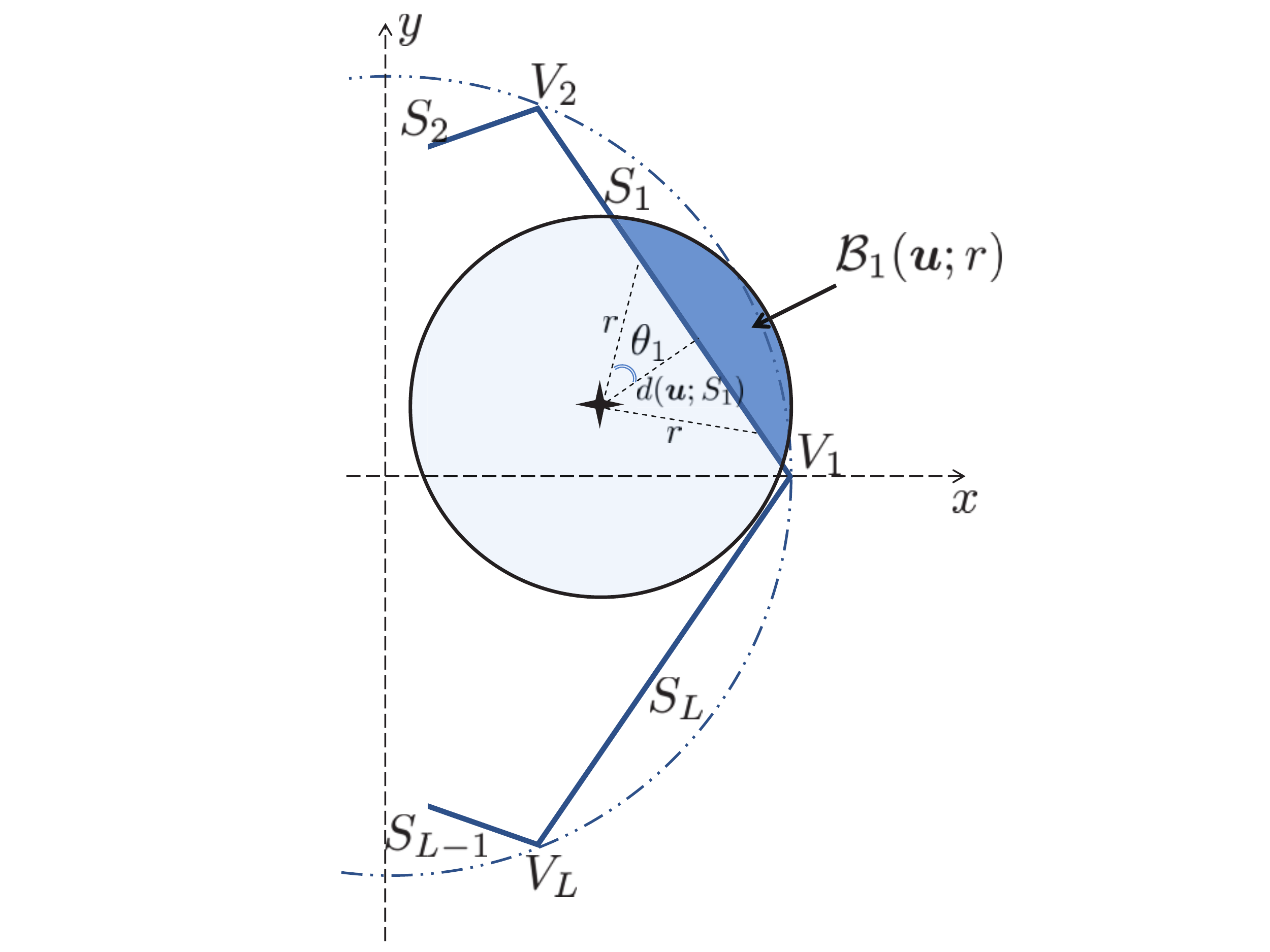}
    }\vspace{-0mm}\hspace{-20mm}
 \subfigure[]
 {
    \includegraphics[width=0.43\textwidth]{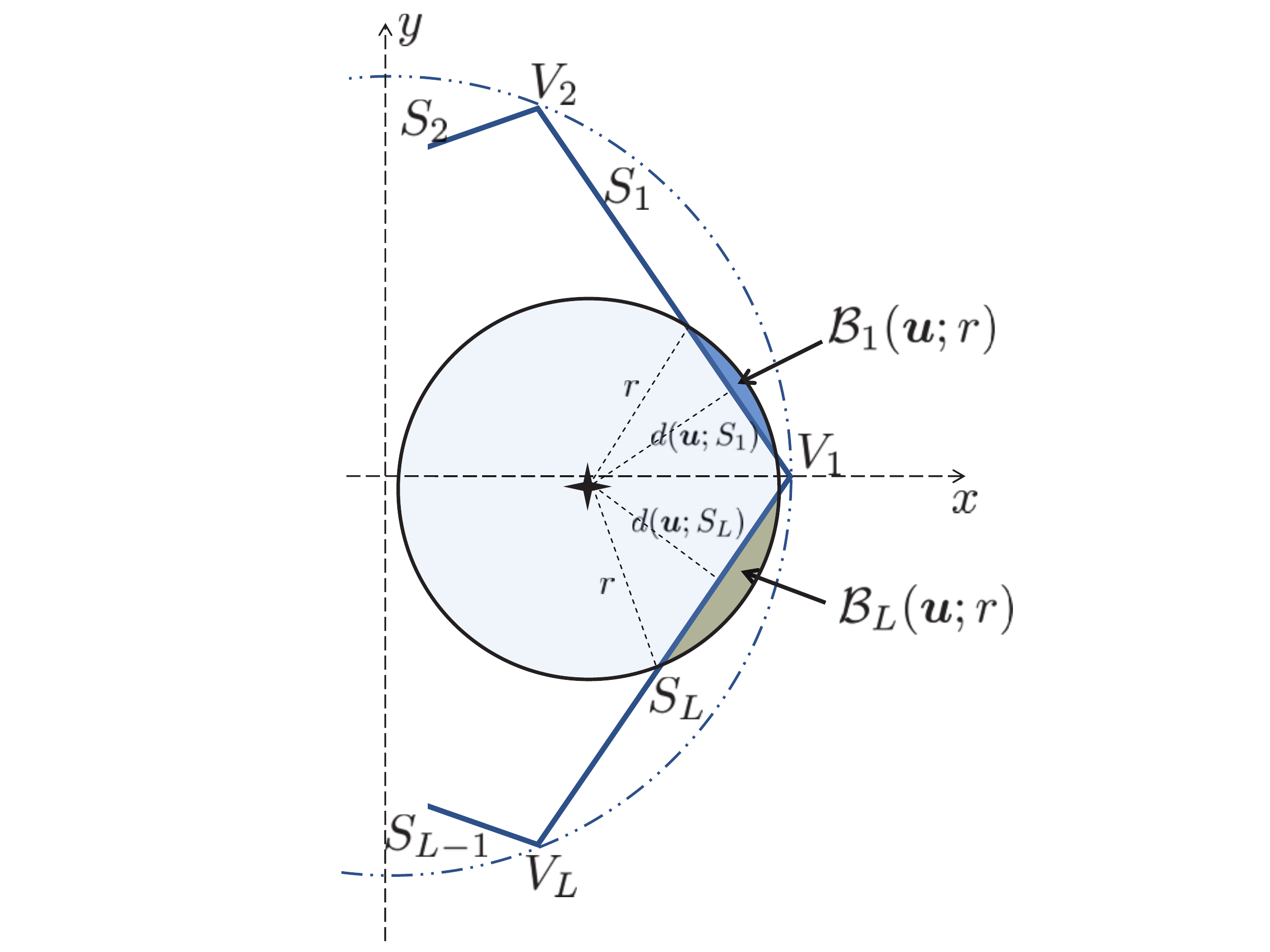}
 }\hspace{-20mm}
 \subfigure[]
 {
    \includegraphics[width=0.43\textwidth]{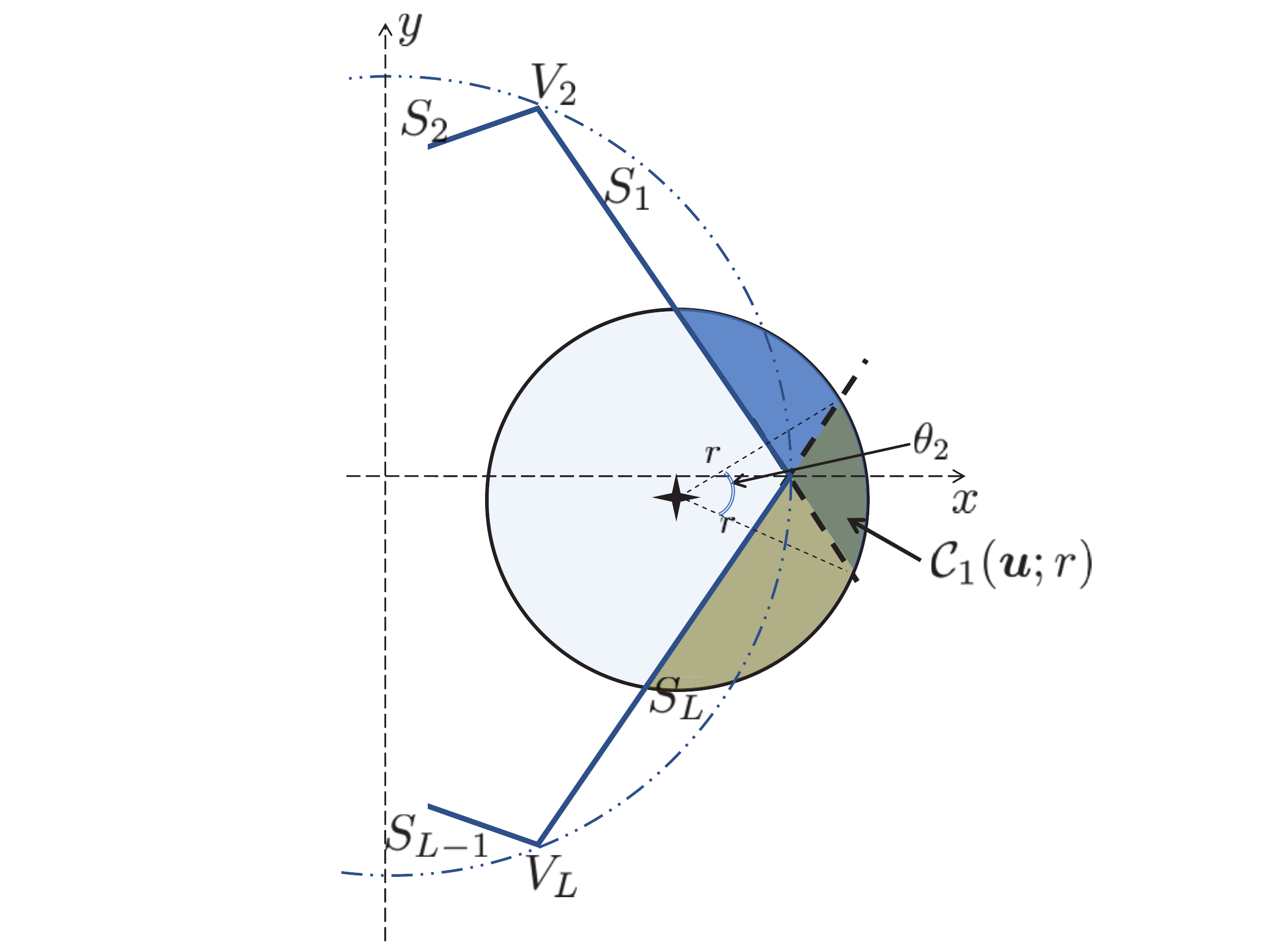}
 }\vspace{-0mm} \hspace{-5mm}
\vspace{-1mm} \caption{Illustration of the effect of sides and
vertices of the polygon on the overlap are: (a) overlap area is
limited by side $S_1$, (b) overlap area is limited by sides $S_1$
and $S_L$ only and (c) overlap area is limited by sides $S_1$ and
$S_L$ with inclusion of the vertex $V_1$.} \label{fig:two}
\end{figure*}
\fi

\subsection{Case 1: Overlap area limited by one side only}
Let $\bbb_1(\dv{u};\rr)$ denote the circular segment formed
outside the side $S_1$, as shown in \figref{fig:two}(a). It is
obvious that the overlap area in this case is given by
$O(\dv{u};\rr)=\pi r^2-B_1(\dv{u};\rr)$, where
$B_1(\dv{u};\rr)=|\bbb_1(\dv{u};\rr)|$. Using polar coordinates, we can find the area $B_1(\dv{u};\rr)$ by
integrating the angle $\theta_1$ (shown in \figref{fig:two}(a)) over $\mathcal{B}_{1}(\dv{u};\rr)$ as
\ifCLASSOPTIONonecolumn
\begin{align}\label{Eq:border_effect_first}
B_1(\dv{u};\rr)\,&=\, \|\bbb_1(\dv{u};\rr) \| \nonumber \\
                 &=\,
                 2\int_{d(\dv{u};S_1)}^{r}\,\rr'\arccos \left(\frac{p(\dv{u};S_1)}{\rr'}\right)d\rr' \nonumber \\
                 &=\,
                 \begin{cases}
            \rr^2\arccos \left(\frac{p(\dv{u};S_1)}{\rr}\right) - (d(\dv{u};S_1))^2\arccos\left( \frac{p(\dv{u};S_1)}{d(\dv{u};S_1)}\right)
            -\\ \quad
                 p(\dv{u};S_1)\bigg(\sqrt{\rr^2-\big(p(\dv{u};S_1)\big)^2} -  \sqrt{(d(\dv{u};S_1))^2-\big(p(\dv{u};S_1)\big)^2}\bigg),   \,\, &  \rr\geq d(\dv{u};S_1);  \\
            0, \,\,&  \textrm{otherwise}; \\
 \end{cases}
\end{align}
\else
\begin{align}\label{Eq:border_effect_first}
B_1(\dv{u};\rr)\,&=\, \|\bbb_1(\dv{u};\rr) \| \nonumber \\
                 &=\,
                 2\int_{d(\dv{u};S_1)}^{r}\,\rr'\arccos \left(\frac{p(\dv{u};S_1)}{\rr'}\right)d\rr' \nonumber
                 \\ \nonumber \\
                 & \hspace{-13mm}=\,
                 \begin{cases}
            \rr^2\arccos \left(\frac{p(\dv{u};S_1)}{\rr}\right) - (d(\dv{u};S_1))^2\arccos\left( \frac{p(\dv{u};S_1)}{d(\dv{u};S_1)}\right)
            -\\ \quad
                 p(\dv{u};S_1)\bigg(\sqrt{\rr^2-\big(p(\dv{u};S_1)\big)^2} - \\ \quad \sqrt{(d(\dv{u};S_1))^2-\big(p(\dv{u};S_1)\big)^2}\bigg),   \,\, & \hspace{-20mm}  \rr\geq d(\dv{u};S_1);  \\
            0, \,\,& \hspace{-20mm}   \textrm{otherwise}; \\
 \end{cases}
\end{align}
\fi

\noindent where $p(\dv{u};S_1)$ and $d(\dv{u};S_1)$ are given in
\eqref{Eq:p_distances_first1} and \eqref{Eq:distances_first1}
respectively.

Generalizing, let $\bbb_\ell(\dv{u};\rr)$ denote the circular
segment which is formed outside the side $S_\ell$. Using
\eqref{Eq:border_effect_first} and the rotation operator in
\eqref{eq:rot}, we
can express $B_\ell(\dv{u};\rr)$ as 
\begin{align}
B_\ell(\dv{u};\rr)\,&=\,  \begin{cases}
            B_1(\rop^{-(\ell-1)}\dv{u};\rr), \,\, &  \rr\geq d(\dv{u};S_\ell);  \\
            0, \,\,&   \textrm{otherwise}. \\
 \end{cases}\label{Eq:border_effect_general1}
\end{align}

\subsection{Case 2: Overlap area limited by two sides with no overlap}
This case is illustrated in \figref{fig:two}(b), where two circular
segments, $\bbb_1(\dv{u};\rr)$ and $\bbb_L(\dv{u};\rr)$, are formed
outside the sides $S_1$ and $S_L$, respectively. Since there is no
overlap between the circular segments, the overlap area is given by
$O(\dv{u};\rr)=\pi r^2-(B_1(\dv{u};\rr)+B_L(\dv{u};\rr))$, where $B_1(\dv{u};\rr)$ is given by \eqref{Eq:border_effect_first} and $B_L(\dv{u};\rr)$ can be found using \eqref{Eq:border_effect_general1}.

\subsection{Case 3: Overlap area limited by two sides with overlap}
This case is illustrated in \figref{fig:two}(c), where there is
overlap between the two circular segments, $\bbb_1(\dv{u};\rr)$ and
$\bbb_L(\dv{u};\rr)$, formed outside sides $S_1$ and $S_L$ due to
the inclusion of the vertex $V_{1}$. Let $\ccc_{1}(\dv{u};\rr) =
\bbb_1(\dv{u};\rr)\cap \bbb_L(\dv{u};\rr)$ denote this overlap
region between the two circular segments. Thus, in this case, the
overlap area is given by $O(\dv{u};\rr)=\pi
r^2-(B_1(\dv{u};\rr)+B_L(\dv{u};\rr)-C_1(\dv{u};\rr))$.

Using polar coordinates, we can find the area $C_1(\dv{u};\rr)$ by
integrating the angle $\theta_2$ (shown in \figref{fig:two}(c)) over $\ccc_{1}(\dv{u};\rr)$ as
\ifCLASSOPTIONonecolumn
\begin{align}\label{Eq:corner_effect_first}
C_1(\dv{u};\rr) &=  \| \ccc_1(\dv{u};\rr)\|\nonumber \\
                &= \int_{d(\dv{u};V_1)}^{\rr}\,\rr' \bigg( \frac{\pi(\el-2)}{\el}  +
                \arccos\left( \frac{p(\dv{u};S_1)}{\rr'}\right) +
                \arccos\left( \frac{p(\dv{u};S_{L})}{\rr'}\right) - \pi \bigg) d\rr' \nonumber \\
                &= \begin{cases}
            \frac{\rr^2}{2} \left(\arccos\left(\frac{p(\dv{u};S_1)}{\rr}\right) \, + \, \arccos\left(\frac{p(\dv{u};S_{L})}{\rr}\right)\right) \, -\,\\ \frac{{(d(\dv{u};V_1))}^2}{2} \left( \arccos\left(\frac{p(\dv{u};S_1)}{d(\dv{u};V_1)}\right) +   \arccos \left(\frac{p(\dv{u};S_{L})}{d(\dv{u};V_1)} \right)\right) \, +\, \\               \frac{p(\dv{u};S_1)}{2} \left(\sqrt{\big(d(\dv{u};V_1)\big)^2-\big(p(\dv{u};S_1)\big)^2} - \sqrt{\rr^2-\big(p(\dv{u};S_1)\big)^2} \right)\,+  \,\\   \frac{p(\dv{u};S_{L})}{2}\left( \sqrt{\big(d(\dv{u};V_1)\big)^2-\big(p(\dv{u};S_{L})\big)^2} -   \sqrt{\rr^2-\big(p(\dv{u};S_{L})\big)^2}   \right)   \, - \\ \frac{\pi}{L}\left( \rr^2 - \big(d(\dv{u};V_1)\big)^2  \right),                   \,\, & \hspace{-40mm}  \rr\geq d(\dv{u};V_1); \\
            0, \,\,& \hspace{-40mm}   \textrm{otherwise}; \end{cases}
\end{align}
\else
\begin{align}\label{Eq:corner_effect_first}
C_1(\dv{u};\rr) &=  \| \ccc_1(\dv{u};\rr)\|\nonumber \\
                &= \int_{d(\dv{u};V_1)}^{\rr}\,\rr' \bigg( \frac{\pi(\el-2)}{\el}
                + \nonumber \\&
                \arccos\left( \frac{p(\dv{u};S_1)}{\rr'}\right) +
                \arccos\left( \frac{p(\dv{u};S_{L})}{\rr'}\right) - \pi \bigg) d\rr' \nonumber
                \\ \nonumber \\
                & \hspace{-10mm}= \begin{cases}
            \frac{\rr^2}{2} \left(\arccos\left(\frac{p(\dv{u};S_1)}{\rr}\right) \, + \, \arccos\left(\frac{p(\dv{u};S_{L})}{\rr}\right)\right) \, -\,\\ \quad \frac{{(d(\dv{u};V_1))}^2}{2} \left( \arccos\left(\frac{p(\dv{u};S_1)}{d(\dv{u};V_1)}\right) +   \arccos \left(\frac{p(\dv{u};S_{L})}{d(\dv{u};V_1)} \right)\right) \, +\, \\ \quad               \frac{p(\dv{u};S_1)}{2} \bigg(\sqrt{\big(d(\dv{u};V_1)\big)^2-\big(p(\dv{u};S_1)\big)^2} - \\ \quad \sqrt{\rr^2-\big(p(\dv{u};S_1)\big)^2} \bigg)\,+  \,\\ \quad   \frac{p(\dv{u};S_{L})}{2}\bigg( \sqrt{\big(d(\dv{u};V_1)\big)^2-\big(p(\dv{u};S_{L})\big)^2} - \\ \quad  \sqrt{\rr^2-\big(p(\dv{u};S_{L})\big)^2}   \bigg)   \, - \\ \quad \frac{\pi}{L}\left( \rr^2 - \big(d(\dv{u};V_1)\big)^2  \right),                   \,\, & \hspace{-40mm}  \rr\geq
            d(\dv{u};V_1);
            \\ \\
            0, \,\,& \hspace{-40mm}   \textrm{otherwise}; \end{cases}
\end{align}
\fi
\noindent \noindent where $d(\dv{u};V_1)$, $p(\dv{u};S_1)$ and $p(\dv{u};S_L)$ are given in
\eqref{Eq:distances_first2}, \eqref{Eq:p_distances_first1} and \eqref{Eq:distances_general1} respectively.

Generalizing, let $\mathcal{C}_\ell(\dv{u};\rr)$ denote the overlap
region between two circular segments adjoining the vertex $V_\ell$
of the polygon. Using \eqref{Eq:corner_effect_first} and the
rotation operator in \eqref{eq:rot}, we can express
$C_\ell(\dv{u};\rr)$ as
\begin{align}
C_\ell(\dv{u};\rr)\,&=\,  \begin{cases}
            C_1(\rop^{-(\ell-1)}\dv{u};\rr), \,\, &  \rr\geq d(\dv{u};V_\ell);  \\
            0, \,\,&   \textrm{otherwise}. \\
 \end{cases}\label{Eq:corner_effect_general1}
\end{align}

\noindent The expressions for the derivatives of $B_\ell(\dv{u};\rr)$ and
$C_\ell(\dv{u};\rr)$, which
are needed in the evaluation of $\frac{\partial
F(\dv{u};\rr)}{\partial\rr}$ in \eqref{Eq:pdf_general}, are provided in \appref{App:derivatives}.

\section{Distance Distributions in Polygons}\label{sec:algo}
In this section, we present our algorithm to use the distance vector in
\eqref{eq:distvector} and the effect of different sides and vertices (\eqref{Eq:border_effect_first}$-$\eqref{Eq:corner_effect_general1})
in order to find the overlap area $O(\dv{u};\rr)$ given any
arbitrary reference point $\dv{u}=[x,\, y]^{T}$ located inside the
$\el$-gon $\mathcal{A}$.

The overall effect of the sides and vertices of the $\el$-gon
depends on the range of the distance $r$ and on the distance between
the reference point and all the sides and vertices. Since an
$\el$-gon has $L$ sides and $L$ vertices, there can be $2L+1$
different ranges for the distance $\rr$. We sort the distance vector
$\mathbf{d}$ in \eqref{eq:distvector} in ascending order and define
$\acute{\mathbf{d}}$ to be the sorted distance vector. Then, the
first range of the distance is $0\leq \rr \leq \acute{d}_{1}$, where
$\acute{d}_{1}$ denotes the first entry of the sorted distance
vector $\acute{\mathbf{d}}$. The next $2L-1$ ranges are
$\acute{d}_{j}\leq\rr\leq \acute{d}_{j+1},\,j=2,\,3,\hdots,\,2L$ and
the last range is $\acute{d}_{2L}\leq\rr$. Thus, in general, we can
write an expression for $O(\dv{u};\rr)$ as
\begin{align}\label{Eq:Reduction Area Cases}
O(\dv{u};\rr) =   \begin{cases}
            O_1(\dv{u};\rr)=\pi r^2, \,\, & 0 \leq \rr \leq \acute{d}_{1};  \\
            O_2(\dv{u};\rr),\,\,&   \acute{d}_{1}\leq\rr\leq\acute{d}_{2}; \\
            \,\,\vdots & \,\,\vdots \\
            O_{2L}(\dv{u};\rr),\,\,&   \acute{d}_{2L-1}\leq\rr\leq\acute{d}_{2L}; \\
            O_{2L+1}(\dv{u};\rr)=A,\,\,&   \rr \geq \acute{d}_{2L}; \\
                    \end{cases}
\end{align}

\noindent where the subscript $j$ in $O_j(\dv{u};\rr)$ denotes the
overlap area for the $j$-th range.

For the first range $0\leq \rr\leq \acute{d}_{1}$,
$S_{1}(\dv{u};\rr)\,=\pi r^2$, since the disk $\cda(\dv{u};\rr)$ will
be completely inside the $\el$-gon. For the last range
$\acute{d}_{2L}\leq\rr$, $O_{2L+1}(\dv{u};\rr)=A$, since the disk
$\cda(\dv{u};\rr)$ will completely overlap with the $\el$-gon
$\mathcal{A}$. For the intermediate ranges, the overlap area may be
limited by any number of sides, with or without overlap between any
two adjacent sides. Also depending on the location of the arbitrary
reference point, some of the distance to the sides and vertices may
be the same. In order to automate the process of finding the unique
set of ranges for the distance $r$ and to calculate the
corresponding overlap area for each unique range, we propose the
Algorithm~\ref{algo:1}.

\begin{algorithm}
\caption{Algorithm to find the overlap area} \label{algo:1}
\begin{algorithmic}

\STATE Step 1: Sort ${\mathbf{d}}$ in \eqref{eq:distvector} in ascending order to obtain $\acute{\mathbf{d}}$%
\STATE Step 2: Determine the index sorting that transforms
${\mathbf{d}}$ into $\acute{\mathbf{d}}$ and obtain the index vector
$\mathbf{k}$

\STATE Step 3: Find the appropriate circular segment areas and the
overlap area

\FOR{each $j$ in $j=1,2,3,\hdots,2L+1$}
\IF{$\acute{d}_{j-1}-\acute{d}_{j}\neq 0$, ($\acute{d}_0=0)$} \STATE
$O_j(\dv{u};\rr) = \pi\rr^2$

\FOR{each $i$ in $i=1,2,3,\hdots,j-1$} \IF{${k}_i\leq\el$} \STATE
$O_j(\dv{u};\rr) = O_j(\dv{u};\rr) - B_{k_i}(\dv{u};\rr)$  \ELSE \STATE
$O_j(\dv{u};\rr) = O_j(\dv{u};\rr) + C_{k_i-\el}(\dv{u};\rr)$ \ENDIF \ENDFOR

 \ENDIF  \ENDFOR
\STATE

\end{algorithmic}
\end{algorithm}

In the proposed algorithm, we sort the distance vector
$\mathbf{d}$ in \eqref{eq:distvector} in ascending order to obtain
the sorted distance vector $\acute{\mathbf{d}}$ and the index vector
$\mathbf{k} = [k_1,\,k_2,\,\hdots,\,k_{2L}]$. The index vector
$\mathbf{k}$ is then used to find the effect of sides and vertices
on the overlap area for each value of range. For each unique range
$\acute{d}_{j-1}-\acute{d}_{j}$ which is determined in the outer \verb"for"
loop, we evaluate the overlap area $O_j(\dv{u};\rr)$ in
\eqref{Eq:Reduction Area Cases} in the inner \verb"for" loop by taking into account the
cumulative effect of all of the sides and vertices. The terms $B_{\el}(\dv{u};\rr)$ and $C_{\el}(\dv{u};\rr)$ are appropriately selected using the index
vector $\mathbf{k}$. Once $O(\dv{u};\rr)$ in \eqref{Eq:Reduction
Area Cases} is computed for each unique range, it can then used to
evaluate \eqref{Eq:CDF_general} and \eqref{Eq:pdf_general}. We have implemented the proposed algorithm in MATLAB\footnote{code available at: \url{http://users.cecs.anu.edu.au/~Salman.Durrani/software.html}}.

\subsection{Algorithm illustration - arbitrary reference point in a square}

In order to illustrate the proposed framework and algorithm, we
consider the problem to find CDF in \eqref{Eq:CDF_general} and the PDF in \eqref{Eq:pdf_general} for a
point $\dv{u}_1 = [R/2,\,-R/2]^T $ located in the middle of the side
$S_4$ of the square~($\el=4$). The distance vector $\mathbf{d}$ in
\eqref{eq:distvector} and the sorted distance vector
$\acute{\mathbf{d}}$ are given by
\begin{align}
\mathbf{d}\,&=\, \left[ \frac{R}{\sqrt{2}},\,  \sqrt{2}R,\,
\frac{R}{\sqrt{2}},\,0,\,
\frac{R}{\sqrt{2}},\,  \frac{\sqrt{10}R}{2},\,     \frac{\sqrt{10}R}{2},\, \frac{R}{\sqrt{2}}    \right], \nonumber \\
\acute{\mathbf{d}}\,&=\, \left[0,\, \frac{R}{\sqrt{2}},\,
\frac{R}{\sqrt{2}},\, \frac{R}{\sqrt{2}},\,  \frac{R}{\sqrt{2}},\,
   \sqrt{2}R,\,     \frac{\sqrt{10}R}{2},\,     \frac{\sqrt{10}R}{2}  \right] \nonumber.
\end{align}

\noindent Thus the index vector ${\mathbf{k}}$ for this
case is given by $\mathbf{k}\,=\,[4,\, 1,\, 2,\,  5,\, 8,\, 3,\, 6,\,7  ]$,
which is employed to determine $O(\dv{u};\rr)$ in
\eqref{Eq:Reduction Area Cases} using the proposed algorithm. Thus the CDF is
\ifCLASSOPTIONonecolumn
\begin{align}\label{Eq:CDF_example}
F(\dv{u}_1;\rr) =   \frac{1}{A}\begin{cases}
            \pi r^2-B_4, \,\, & 0 \leq \rr \leq R/\sqrt{2};  \\
            \pi r^2-(B_1+B_2+B_4-C_1-C_4), \,\,&   R/\sqrt{2} \leq\rr\leq \sqrt{2}R; \\
             \pi r^2-(B_1+B_2+B_3+B_4-C_1-C_4),\,\,&   \sqrt{2}R \leq\rr\leq \sqrt{10}R/2; \\
            A,\,\,&  \rr \geq \sqrt{10}R/2. \\
                    \end{cases}
\end{align}
\else
\begin{align}\label{Eq:CDF_example}
&F(\dv{u}_1;\rr) =   \frac{1}{A} \times \nonumber \\& \begin{cases}
            \pi r^2-B_4, \,\, & 0 \leq \rr \leq \frac{R}{\sqrt{2}};  \\
            \pi r^2-(B_1+B_2+B_4-C_1-C_4), \,\,&    \frac{R}{\sqrt{2}}\leq\rr\leq \sqrt{2}R; \\
             \pi r^2-(B_1+B_2+B_3+B_4-C_1-C_4),\,\,&   \sqrt{2}R \leq\rr\leq \frac{\sqrt{10}R}{2}; \\
            A,\,\,&  \rr \geq \frac{\sqrt{10}R}{2}. \\
                    \end{cases}
\end{align}
\fi
\noindent Substituting \eqref{Eq:CDF_example}, and its derivative which is obtained using
the expressions in \appref{App:derivatives}, in
\eqref{Eq:pdf_general}, we obtain the PDF of the distance to the
$n$-th neighbour, which is shown in \figref{fig:square_plot} for
$R=1$, $N=5$ and $n=1,\,2,\hdots,\,5$. We can see that the
simulation results, which are averaged over 10,000 simulation runs,
match perfectly with the analytical results, verifying the proposed
framework and algorithm.
%
\ifCLASSOPTIONonecolumn
\begin{figure*}[t]
 \centering
 \hspace{-18mm}
    \includegraphics[width=0.8\textwidth]{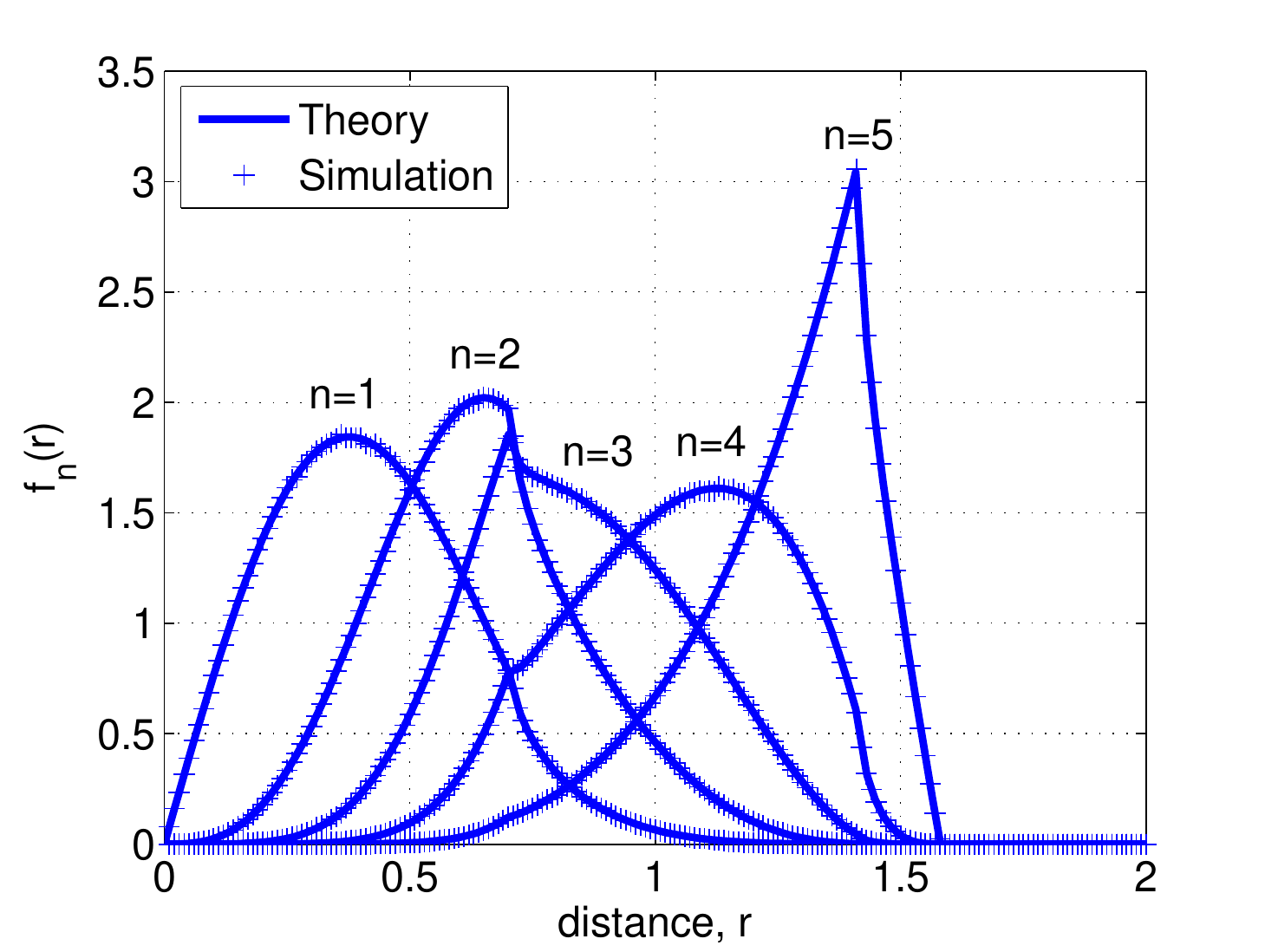}
    \vspace{-0mm} \hspace{-15mm}

\vspace{-1mm} \caption{The PDF $f_n(\dv{u};\rr)$ of the distance to
$n$-th nearest neighbour for a point located midway between the
$V_1$ and $V_4$ on the side $S_4$ of the square~($L=4$). The
circumradius $R=1$ and the number of nodes is $N=5$. }
\label{fig:square_plot}
\end{figure*}
\else
\begin{figure}[t]
 \centering
 \hspace{-10mm}
    \includegraphics[width=0.48\textwidth]{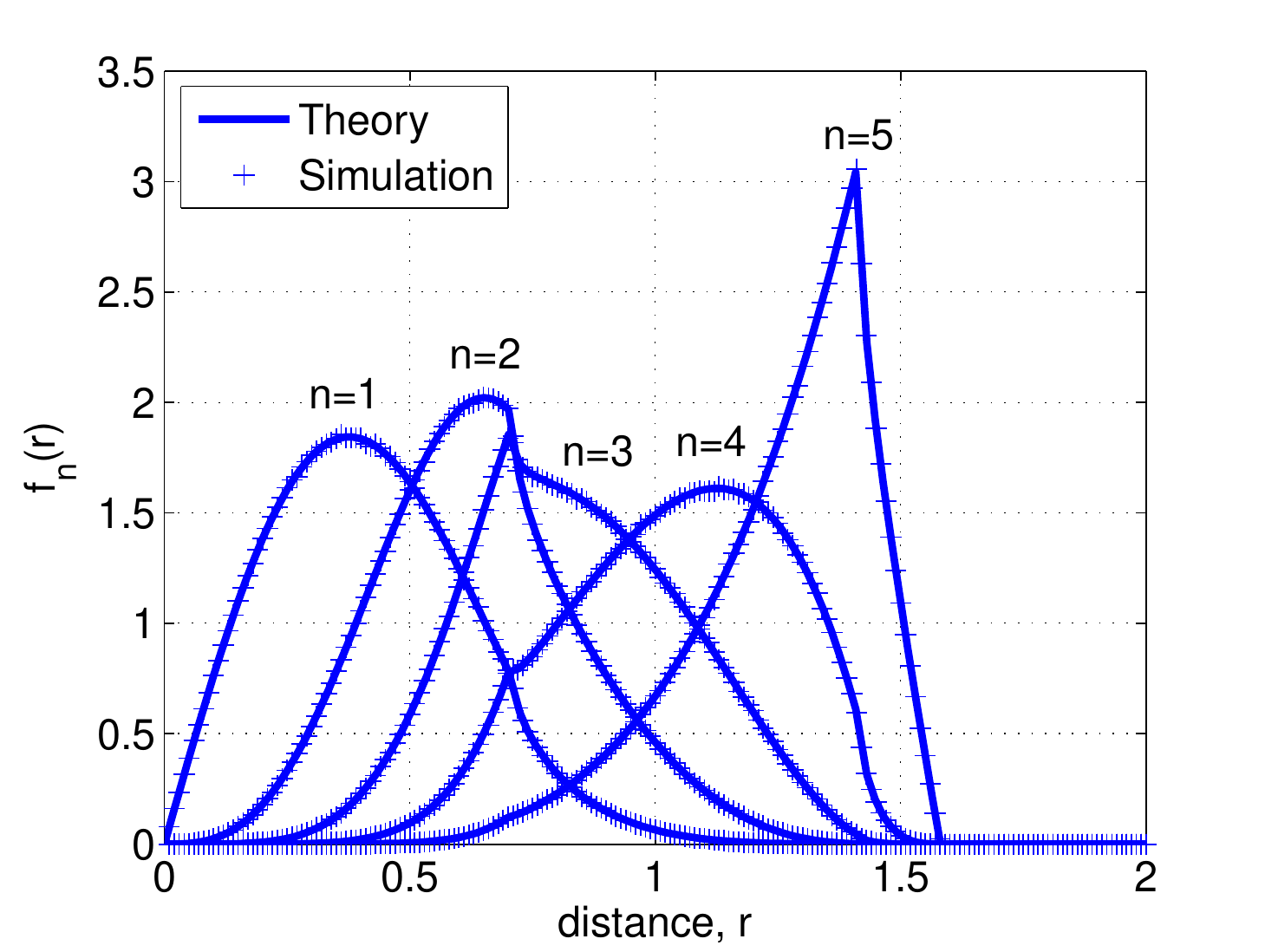}
    \vspace{-0mm} \hspace{-15mm}

\vspace{-1mm} \caption{The PDF $f_n(\dv{u};\rr)$ of the distance to
$n$-th nearest neighbour for a point located midway between the
$V_1$ and $V_4$ on the side $S_4$ of the square~($L=4$). The
circumradius $R=1$ and the number of nodes is $N=5$. }
\label{fig:square_plot}
\end{figure}
\fi

\subsection{Special case - reference point at the center of the polygon}
Consider the special case that the reference point is located
at the center~$O$ of the polygon, i.e., at the origin
$\dv{u}_2=[0 \; 0]^T$. All of the sides and also all of the vertices are
equidistant from the center of the polygon and the rotation operator
does not affect the point located at the origin. This implies that $d(\dv{u}_2;S_\ell)= d(\dv{u}_2;S_1)=R\tan(\theta/2)$, $d(\dv{u}_2;V_\ell)=d(\dv{u}_2;V_1)=R$, $B_\ell= B_1$ and $C_\ell=C_1$ for $\ell=2,3,\hdots,L$. By employing these relations and the proposed
algorithm, we obtain the CDF expression for the following three possible ranges
\begin{align}\label{Eq:cdf_center}
F(\dv{u}_2;\rr)\,=\,   \frac{1}{A}\begin{cases}
            {\pi\rr^2}, \,\, & 0 \leq \rr \leq R\tan\left(\frac{\theta}{2}\right);  \\
            {\pi\rr^2}-\,L B_1, \,\,   &   R\tan\left(\frac{\theta}{2}\right)\leq\rr\leq R; \\
            A,\,\,&  \rr \geq R; \\
                    \end{cases}
\end{align}

\noindent where
$B_1 =  \rr^2\arccos
\big(\frac{R}{\rr}\tan\left(\frac{\theta}{2}\right)\big) -
R\tan\frac{\theta}{2}\sqrt{\rr^2-  R^2 \tan^2\left(\frac{\theta}{2}\right)}
$ and $\theta$ is defined in \eqref{eq:angles}. Substituting \eqref{Eq:cdf_center} in \eqref{Eq:pdf_general} reproduces
the result in~\cite{Srinivasa-2010}.

\subsection{Special case - reference point at the vertex of the
polygon}

Consider the special case when the arbitrary point is located
at one of the vertex of the polygon. Let the point $\dv{u}_3$ be
located at $V_1$, that is $\dv{u}_3 = [R \; 0]^T$. From the vertex
$V_1$, the sides $S_\ell$, $S_{L+1-\ell}$ and the vertices
$V_{\ell}$, $V_{L+2-\ell}$ are \emph{all} equidistant for
$\ell=2,3,\hdots\lfloor L/2 \rfloor$, where $\lfloor \cdot\rfloor$
denotes integer floor function. Hence, there are
$\frac{L}{2}+1$ possible ranges of the distance. By using the
proposed algorithm to determine the overlap area $O(\dv{u};\rr)$, we
obtain the CDF expression for these ranges as
\ifCLASSOPTIONonecolumn
\begin{align}\label{Eq:distances_corner_special case}
F(\dv{u}_3;\rr)\,=\,  \frac{1}{A}\begin{cases}
            \pi\rr^2 - \sum\limits_{i=1}^{\ell}(B_i + B_{L+i-1}-C_i -C_{L+2-i} ),   & \\ \quad\quad\quad\quad\quad\quad \quad   d(\dv{u}_3,V_{\ell}) \leq \rr
            \leq  d(\dv{u}_3,V_{\ell+1}),\quad \ell=1,2,\hdots \big\lfloor\frac{L}{2}\big\rfloor,\ C_{L+1}=0;&        \\
            A,\,\,& \hspace{-103mm} d(\dv{u}_3,V_{\lfloor \frac{L}{2} \rfloor +1})
            \leq\rr; \\
                    \end{cases}
\end{align}
\else \begin{align}\label{Eq:distances_corner_special case}
F(\dv{u}_3;\rr)\,=\,  \frac{1}{A}\begin{cases}
            \pi\rr^2 - \sum\limits_{i=1}^{\ell}(B_i + B_{L+i-1}-C_i -C_{L+2-i} ),   & \\ \quad\quad\quad\quad   d(\dv{u}_3,V_{\ell}) \leq \rr
            \leq  d(\dv{u}_3,V_{\ell+1}), \nonumber & \\ \quad\quad\quad\quad \ell=1,2,\hdots \big\lfloor\frac{L}{2}\big\rfloor,\ C_{L+1}=0;&       \\ \\
            A,\,\,& \hspace{-52mm} d(\dv{u}_3,V_{\lfloor \frac{L}{2} \rfloor +1})
            \leq\rr; \\
                    \end{cases}
\end{align}
\fi
which is then used to the evaluate the PDF of the
distance to the $n$-th neighbour by employing
\eqref{Eq:pdf_general}. \figref{fig:corner_plot} shows the plot of
the PDF of the distance to the farthest neighbour ($n=10$) from the
vertex $V_1$, with $N=10$ nodes distributed inside a $\el$-gon with
area $A=100$, for $\el =3,4,6$ and $\el=\infty$ which corresponds to
a disk. For a disk, we can easily use the overlap area approach to obtain the CDF as
\ifCLASSOPTIONonecolumn
\begin{equation}\label{Eq:cdf_circle}
F(\dv{u};\rr) = \frac{1}{\pi R^2}\begin{cases} \pi\rr^2, &  0
\leq \rr \leq R-\psi(\dv{u}); \\
r^2\arccos\left(\frac{r^2+\psi^2(\dv{u})-R^2}{2 r\,
\psi(\dv{u})}\right)+R^2\arccos \left(\frac{R^2+\psi^2(\dv{u})-r^2}{2 R
\,\psi(\dv{u})}\right)-\frac{\sqrt{\zeta}}{2}, &  R-\psi(\dv{u})< \rr
\leq R+\psi(\dv{u});
\\
\pi R^2, &  R+d<\rr;
\end{cases}
\end{equation}
\else
\begin{eqnarray}\label{Eq:cdf_circle}
&& F(\dv{u};\rr) = \frac{1}{\pi R^2} \times  \nonumber \\ && \begin{cases} \pi\rr^2, &  0\leq \rr \leq R-\psi(\dv{u}); \\
r^2\arccos\left(\frac{r^2+\psi^2(\dv{u})-R^2}{2 r\,\psi(\dv{u})}\right) & \\
+R^2\arccos \left(\frac{R^2+\psi^2(\dv{u})-r^2}{2 R\,\psi(\dv{u})}\right) &  \\
-\frac{\sqrt{\zeta}}{2},& R-\psi(\dv{u})\leq \rr\leq R+\psi(\dv{u});
\\
\pi R^2, &  R+d<\rr;
\end{cases}
\end{eqnarray}
\fi


\noindent where $\zeta =
(\psi(\dv{u})+r+R)(-\psi(\dv{u})+r+R)(\psi(\dv{u})-r+R)(r+\psi(\dv{u})-R)$
and $\psi(\dv{u}) = \sqrt{x^2+y^2}$ denotes the distance of the
point $\dv{u}$ from the origin. Note that \eqref{Eq:cdf_circle} is similar to~\cite[eq. (11)]{Bettstetter:2004} but with different range conditions. Setting $\psi(\dv{u}_3)= R$ and substituting \eqref{Eq:cdf_circle} in \eqref{Eq:pdf_general} produces the result for a disk shown in \figref{fig:corner_plot}.
For the simulation results, we have used the acceptance-rejection method~\cite{Martinez-2001} to uniformly distribute the points inside the
$\el$-gon and averaged the results over 10,000 simulation runs.
Again it can be seen that the simulation results match perfectly
with the analytical results.
\ifCLASSOPTIONonecolumn
\begin{figure*}[t]
 \centering
 \hspace{-18mm}
    \includegraphics[width=0.8\textwidth]{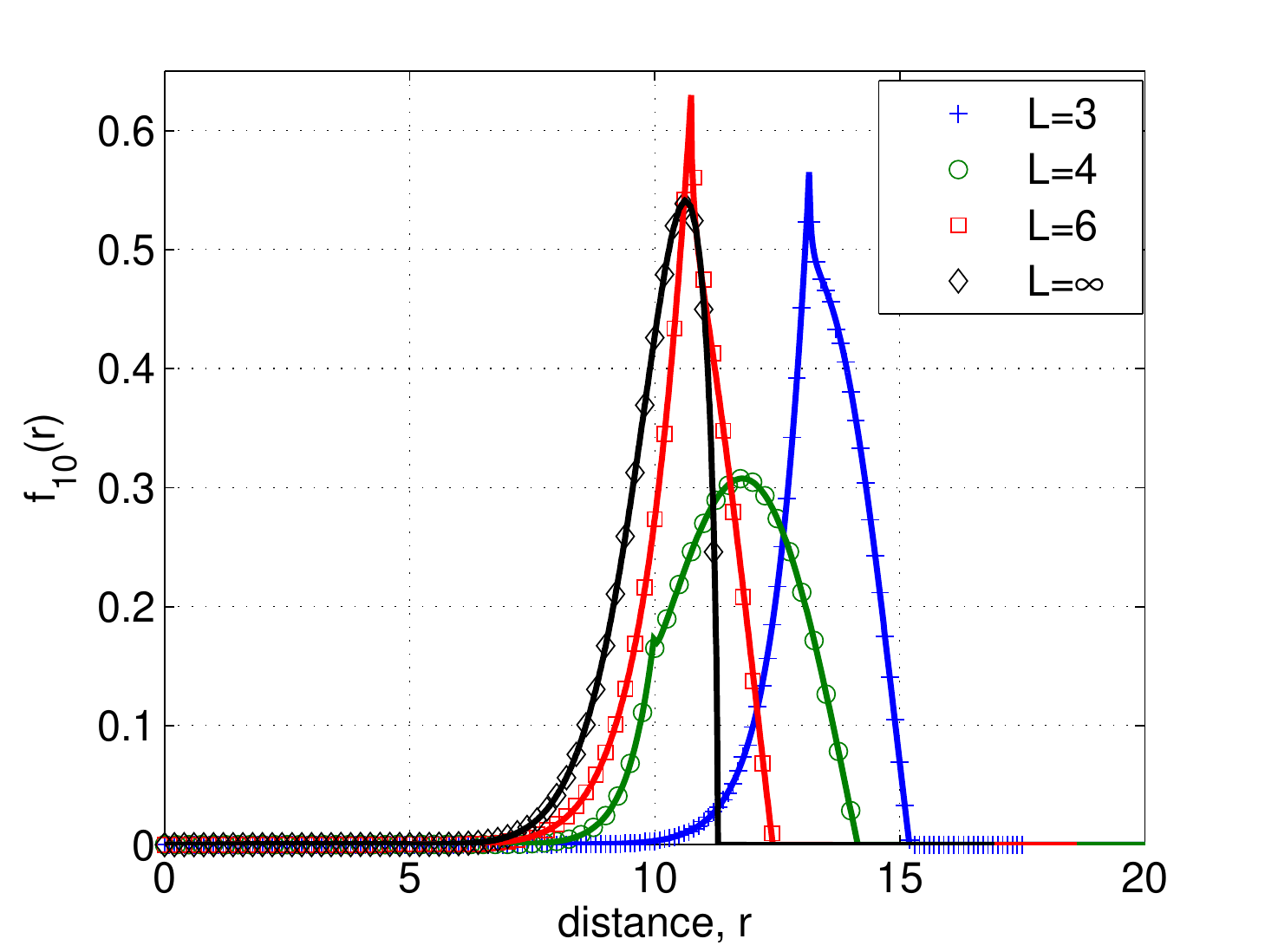}
    \vspace{-0mm} \hspace{-15mm}

\vspace{-1mm} \caption{The PDF $f_{10}(\dv{u};\rr)$ of the distance
to farthest neighbour for a point located at the vertex of the
polygon with area $A=100$ and $N=10$ nodes, for triangle~($L=3$),
square~($L=4$), hexagon~($L=6$) and disk~($L=\infty$). Solid lines
show analytical result using our proposed algorithm and markers
provide verification using simulation.}
\label{fig:corner_plot}
\end{figure*}
\else
\begin{figure}[t]
 \centering
 \hspace{-18mm}
    \includegraphics[width=0.48\textwidth]{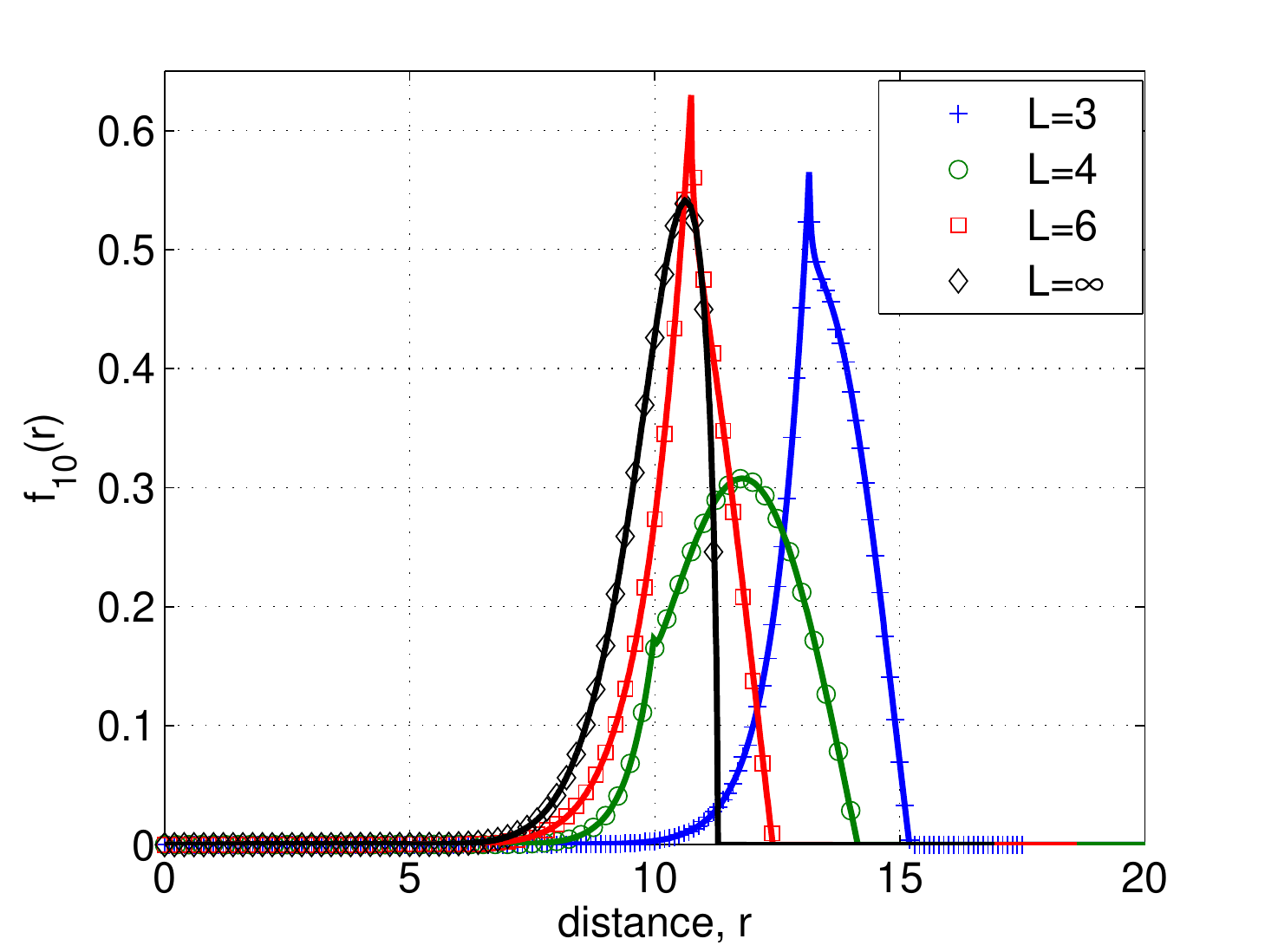}
    \vspace{-0mm} \hspace{-15mm}

\vspace{-1mm} \caption{The PDF $f_{10}(\dv{u};\rr)$ of the distance
to farthest neighbour for a point located at the vertex of the
polygon with area $A=100$ and $N=10$ nodes, for triangle~($L=3$),
square~($L=4$), hexagon~($L=6$) and disk~($L=\infty$). Solid lines
show analytical result using our proposed algorithm and markers
provide verification using simulation.} \label{fig:corner_plot}
\end{figure}
\fi

\section{Conclusions}

In this paper, we have derived the exact cumulative density function of the distance between a randomly located node and any arbitrary reference point inside a regular $\el$-sided polygon. We have used it to obtain the closed-form PDF of the Euclidean distance between any arbitrary reference point and its $n$-th neighbour node, when $N$ nodes are uniformly and independently distributed inside a regular $\el$-sided polygon. We have provided examples to demonstrate the generality of our proposed framework. Future work can exploit the knowledge of these general distance distributions to model and analyse the wireless network characteristics, such as connectivity~\cite{Zubair-2013}, interference and outage probability, from the perspective of an arbitrary node located anywhere (i.e. not just the center) in the finite coverage area.

\begin{appendices}
\section{Derivatives}\label{App:derivatives}
By employing the Leibniz integral rule, we can express the
derivatives of $B_\ell(\dv{u};\rr)$ and $C_\ell(\dv{u};\rr)$, which
are required in the evaluation of $\frac{\partial F(\dv{u};\rr)}{\partial\rr}$ in
\eqref{Eq:pdf_general}, as
\ifCLASSOPTIONonecolumn
\begin{align}
\frac{\partial
B_\ell(\dv{u};\rr)}{\partial\rr}\,&=\,2\rr\arccos \left(\frac{p(\dv{u};S_\ell)}{\rr}\right)\\
\frac{\partial
C_\ell(\dv{u};\rr)}{\partial\rr}\,&=\, \rr \bigg(
                \arccos \left(\frac{p(\dv{u};S_\ell)}{\rr}\right) +
                \arccos \left(\frac{p(\dv{u};S_{\ell-1})}{\rr}\right) - \frac{2\pi}{L} \bigg)
\end{align}
\else
\begin{align}
\frac{\partial
B_\ell(\dv{u};\rr)}{\partial\rr}\,&=\,2\rr\arccos \left(\frac{d(\dv{u};S_\ell)}{\rr}\right)\\
\frac{\partial C_\ell(\dv{u};\rr)}{\partial\rr}\,&=\, \rr \bigg(
                \arccos \left(\frac{p(\dv{u};S_\ell)}{\rr}\right) +
                \nonumber \\ &
                \arccos \left(\frac{p(\dv{u};S_{\ell-1})}{\rr}\right) - \frac{2\pi}{L} \bigg)
\end{align}
\fi
\end{appendices}

\ifCLASSOPTIONonecolumn
\vspace{-0.25in}
\enlargethispage{0.25 in}
\fi

\end{document}